\documentclass[11pt]{article}

\usepackage{amssymb}
\usepackage[numbers,sort&compress]{natbib}
\usepackage{amsmath,amsfonts,epsfig}
\usepackage{bm}
\usepackage{simplewick}
\usepackage[dvipsnames]{xcolor}
\usepackage{latexsym}
\usepackage{mathrsfs}
\usepackage{braket}
\usepackage{array}
\usepackage{float}
\usepackage{verbatim,graphicx}
\usepackage[%
             colorlinks=true,urlcolor=MidnightBlue,linkcolor=MidnightBlue,citecolor=OliveGreen,
             pdfpagelabels=true,hypertexnames=true,
            plainpages=false,naturalnames=false,
             ]{hyperref}
\usepackage{subcaption}
\def\bea{\begin{eqnarray}}
\def\eea{\end{eqnarray}}
\def\be{\begin{equation}}
\def\ee{\end{equation}}
\def\ba{\begin{array}}
\def\ea{\end{array}}

\def\k{{\bf k}}

\def\R{{\mathcal R}}

\def\P{{\mathcal P}}

\begin{document}

\setlength\arraycolsep{2pt}

\renewcommand{\theequation}{\arabic{section}.\arabic{equation}}
\setcounter{page}{1}

\begin{titlepage}

\begin{center}

\vskip 1.0 cm

{\Huge Scale invariance of the primordial tensor power spectrum}
\\

\vskip 1.0cm

{\large
Gonzalo A. Palma, Basti\'an Pradenas, Walter Riquelme\\ and Spyros Sypsas
}

\vskip 0.5cm

{\it
Grupo de Cosmolog\'ia y Astrof\'isica Te\'orica, Departamento de F\'{i}sica, FCFM, \mbox{Universidad de Chile}, Blanco Encalada 2008, Santiago, Chile.
}

\vskip 1.5cm

\end{center}

\begin{abstract} 

Future cosmic microwave background polarization experiments will search for evidence of primordial tensor modes at large angular scales, in the multipole range $4 \leq \ell \leq 50$. Because in that range there is some mild evidence of departures from scale invariance in the power spectrum of primordial curvature perturbations, one may wonder about the possibility of similar deviations appearing in the primordial power spectrum of tensor modes. Here we address this issue and analyze the possible presence of features in the tensor spectrum resulting from the dynamics of primordial fluctuations during inflation. We derive a general, model independent, relation linking features in the spectra of curvature and tensor perturbations. We conclude that even with large deviations from scale invariance in the curvature power spectrum, the tensor spectrum remains scale invariant for all observational purposes.

\end{abstract}

\end{titlepage}

\setcounter{equation}{0}
\section{Introduction}

The simplest models of cosmic inflation~\cite{Guth:1980zm,Linde:1981mu, Starobinsky:1980te,Albrecht:1982wi} predict both scalar and tensor primordial fluctuations, characterized by a set of nearly scale invariant power spectra. While cosmic microwave background (CMB) observations have enabled us to tightly constrain the power spectrum of scalar perturbations, a detection of primordial gravity waves (in the form of B-modes) remains a pending challenge. Current efforts to observe the CMB polarization will reach the limits of cosmic variance, allowing us to either measure or constrain the tensor-to-scalar ratio $r$ down to $r \sim 0.01$ - $0.002$~\cite{Harrington:2016jrz, Abazajian:2016yjj,Suzuki:2015zzg,Ahmed:2014ixy}. The observation of B-modes in the CMB would give us access to the value of the Hubble expansion rate $H$ during inflation, reinforcing the idea that the Hot Big Bang era was preceded by a stage of dramatic accelerated expansion.

Although current CMB observations are compatible with a nearly scale invariant power spectrum for curvature perturbations~\cite{Aghanim:2015xee}, there are some hints of scale dependent features present in the spectrum at certain multipoles~\cite{Hazra:2013nca,Hunt:2013bha,Hazra:2014jwa,Hunt:2015iua}. The shape and size of such features could in principle allow us to discriminate the type of physics that played a role during inflation, since their appearance in the primordial spectra would invalidate the simplest models of inflation, forcing us to consider models in which non-trivial degrees of freedom interacted with primordial curvature fluctuations around horizon crossing~\cite{Aghanim:2015xee,Bennett:1996ce,Hinshaw:2003ex,Spergel:2003cb,Peiris:2003ff,Ade:2013zuv,Ade:2015xua,Ade:2013kta,Benetti:2016tvm,Novaes:2015uza,Benetti:2013cja,Gariazzo:2016blm,Ashoorioon:2008qr,Gariazzo:2015qea,Gao:2015aba,Cai:2015xla,GallegoCadavid:2016wcz,Hazra:2016fkm} (see also~\cite{Polarski:1995zn,Lesgourgues:1998mq,Polarski:1999fb} for early work on features of the tensor spectrum and~\cite{Chluba:2015bqa} for an up-to-date review). The prospects of unveiling physics beyond the single-field slow-roll paradigm has also propelled new ideas to analyze the presence of such features in 21 cm and Large Scale Structure observations~\cite{Chen:2016zuu,Xu:2016kwz,Chen:2016vvw,Ballardini:2016hpi}. 

The effective field theory (EFT) approach to inflation~\cite{Cheung:2007st,Weinberg:2008hq} is particularly useful to understand the appearance of features in the primordial spectra. This formalism allows one to study models of inflation beyond the canonical single field paradigm by incorporating the sound speed at which curvature fluctuations propagate, as a parameter in the Lagrangian for perturbations. Within this framework, features are the consequence of time variations of background quantities appearing in the Lagrangian describing the dynamics of the lowest energy fluctuations. These time variations break -- in a controlled way -- the standard behavior required in single field slow-roll inflation, producing localized features in the spectra, though without invalidating inflation as a mechanism to explain the origin of primordial fluctuations in a way compatible with observations. Given that the source of features may be traced back to background parameters that affect the evolution of all perturbations, features appearing in different $n$-point correlation functions would be necessarily correlated~\cite{Achucarro:2014msa,Achucarro:2013cva,Achucarro:2012fd,Gong:2014spa,Fergusson:2014tza,Palma:2014hra,Mooij:2015cxa,Cadavid:2015iya,Torrado:2016sls,Meerburg:2015owa,Appleby:2015bpw,Mooij:2016dsi}. In the case of scalar perturbations, a powerful way to study such time-dependent departures from slow-roll is the joint estimator analysis of two- and three-point correlation functions~\cite{Meerburg:2015owa}, since a detection of correlated signals in the power spectrum and bispectrum would increase the statistical significance of these features.

In this article we explore the possibility of establishing a novel class of cross correlation between spectra. Specifically, the questions we wish to address are the following: \emph{If features in the primordial scalar power spectrum are confirmed, would they also show up in the tensor power spectrum? In addition, if the scale suppression of the angular power spectrum in the multipole range $4 \leq \ell \leq 50$ is found to be of primordial origin, what type of signal should we expect in the angular power spectrum of B-modes?} To that end, we study the effect of time dependent backgrounds on the dynamics of fluctuations in order to correlate features in the power spectra of scalar and tensor modes. Our main result is that features $\Delta \P_T / \P_{T}$, appearing in the tensor power spectrum $\P_{T}$, are correlated to features $\Delta \P_S / \P_S$, appearing in the scalar spectrum $\P_S$ in Fourier space, in the following way
\be \label{eq:cor-in}
\frac{d^2}{d\ln k^2} \left( \frac{\Delta \P_T}{\P_{T}} \right)= 6 \epsilon_0  \frac{\Delta \P_S}{\P_{S}},
\ee
where $\epsilon_0$ is the (constant) average value of the slow-roll parameter $\epsilon = - \dot H/H^2$. This expression tells us that any feature appearing in the tensor spectrum is in general suppressed with respect to those appearing in the scalar spectrum~\cite{Hu:2014hoa}. This suppression is two-fold: On the one hand, $\epsilon_0$ must be small in order to keep inflation valid as a mechanism to produce fluctuations over a large range of scales. On the other hand, the $\ln k$-derivatives must be large in order for features to be observable in the scalar power spectrum.\footnote{As we shall see in the next section, observable features in the spectra must have an identifiable structure over a range of scales smaller than $\ln k$. This implies that $\ln k$-derivatives acting on either $\Delta \mathcal P_T$ or $\Delta \mathcal P_S$ must be large.} Note that this approach is model independent since it takes the scalar power spectrum data as an input without reference to the mechanism that produces the features.

Our results show that any strong departure of scale invariance in the scalar spectrum must come together with a consequential departure in the tensor spectrum, but at a level that is too small to be observed. As a corollary, any future observation of scale invariance departures in the tensor spectrum cannot be of primordial origin, unless some exotic mechanism underlies their origin. For example, models where the only background quantity experiencing rapid variations is the tensor sound speed will have features only in the tensor spectrum~\cite{Cai:2015yza}. On the other hand, non Bunch-Davies initial conditions may lead to features in the two spectra with the same amplitude~\cite{Broy:2016zik}. In this work, however, we are interested in predicting the scale dependence of the tensor spectrum from the scalar power spectrum, highlighting the perspective of a joint analysis of the two spectra. Having this in mind, in the particular case of the observed deficit of the angular power spectrum around $\ell\sim 20$, we conclude that coming CMB polarization experiments should not encounter any scale dependence of the spectrum around that region.

The article is organized as follows: In Section~\ref{sec:correlation} we present the method used and derive the correlation of the two power spectra for the cases where {\it i}) features appear due to sudden variations of the Hubble scale, and {\it ii}) variations in both the Hubble scale and the sound speed are responsible for features. In Section~\ref{sec:examples}, we present results for the tensor power spectrum in the low $\ell$ region, modeling the features in the scalar signal with a Gaussian and a cosine function. Finally, we conclude in Section~\ref{sec:conclusions}.

\section{Correlation of power spectra}\label{sec:correlation}

In this section we apply the methods elaborated in~\cite{Achucarro:2012fd,Palma:2014hra} to correlate features appearing in the tensor and scalar power spectra. Our method is based on the \emph{in-in} formalism to study the evolution of quantum fluctuations on a time dependent quasi-de Sitter background~\cite{Maldacena:2002vr,Weinberg:2005vy}. Another widely used method to study features is the so called generalized slow-roll formalism~\cite{Stewart:2001cd,Choe:2004zg,Adshead:2011bw,Dvorkin:2009ne}.

\subsection{Preliminaries}

Let us set the ground for the computation by first writing down the quadratic actions for the scalar and tensor perturbations in Fourier space. For the scalar part we will consider the primordial curvature perturbation $\R$ in comoving gauge. On the other hand, for the tensor part we will work with the traceless and transverse perturbation $\gamma_{ij}$ as:
\begin{equation}
\gamma_{ij} (\k , \tau) \equiv h_{+} (\k , \tau) e_{ij}^+ (\k) + h_{\times} (\k , \tau) e_{ij}^\times (\k) ,
\end{equation}
where $\k$ is the wave vector (or momenta), and $e_{ij}^+ (\k) $ and $e_{ij}^\times (\k) $ are the elements of a time independent basis for tensors satisfying $\delta^{ij} e_{ij} = 0$ and $k^i e_{ij} = 0$. We may further define canonically normalized fields $u$ and $f_{+,\times}$ as
\begin{equation}
u=z\mathcal{R},\qquad f_{+,\times} =a(t) h_{+,\times}, \qquad z\equiv\sqrt{2\epsilon} \frac{a}{c_s},
\end{equation}
where $a(t)$ is the scale factor, $c_s$ is the sound speed of the curvature perturbations and $\epsilon=-\dot H/H^2$ the first Hubble slow-roll parameter. In these variables, the quadratic actions for scalar and tensor modes in conformal time $\tau$ are found to be
\begin{eqnarray} \label{S_s}
S^{(2)}_{S} &=& \frac{1}{2}\int d\tau \, d^3 k \left[ (u')^2 + c_{s}^2 k^2 u^2+\frac{z''}{z}u^2\right], \\ \label{S_t}
S^{(2)}_{T} &=& \frac{1}{2}\int d\tau \, d^3k \left[(f')^2 + k^2  f^2+\frac{a''}{a}f^2  \right],
\end{eqnarray}
where we have chosen units such that $m_{\rm Pl} =1$, while keeping only one polarization mode for simplicity. Notice that primes ($'$) represent derivatives with respect to $\tau$. The background quantities $z''/z$ and $a''/a$ may be written as
\bea
\frac{z''}{z} &=& (aH)^2\left( 2-\epsilon +\frac{1}{2}\eta -s \right) \left( 1 + \frac{1}{2} \eta - s \right) + aH \left( \frac{\eta'}{2} -s' \right) , \label{zppz} \\
\frac{a''}{a} &=& (aH)^2\left(2 - \epsilon\right),
\eea
where $\eta=\epsilon'/\epsilon aH$ and $s= c_s'/c_s aH$.

\subsection{Rapidly time varying backgrounds}

To describe the origin of features, we may split each action into a zeroth order term, that describes the evolution of fluctuations in a quasi-de Sitter spacetime, and an interaction term, that contains the rapidly varying contributions of the background. To do so, we will assume that the background is such that $\epsilon$ remains small ($\epsilon \ll 1$) throughout the whole relevant period where features are sourced. To model this behavior we will take $\epsilon$ to be of the form:
\be
\epsilon = \epsilon_0 + \Delta \epsilon  ,  \qquad   |\Delta \epsilon |  \ll  \epsilon_0  ,  \label{epsilon-small} 
\ee
where $\epsilon_0$ is (for any practical purpose) a constant, and $\Delta \epsilon (\tau)$ contains information about the sudden variations of the background. One could consider that $\epsilon_0 =-\dot H_0/H_0^2$, where $H_0$ is the slowly varying part of the Hubble expansion rate.  In the same manner, $\eta$ will have two contributions:
\be
\eta = \eta_0 + \Delta \eta ,  \qquad  \Delta \eta   = -\frac{1}{\epsilon_0}\tau \Delta \epsilon',
\ee
where $\eta_0 = - \dot \epsilon_0 / H_0 \epsilon_0$. Given that we are taking $\epsilon_0$ as a slowly varying function, we may neglect $\eta_0$ against $\Delta \eta$ and simply take
\be \label{eta-def}
\eta  = -\frac{1}{\epsilon_0}\tau \Delta \epsilon' .
\ee
We will additionally assume that $\eta$ remains small at all times:
\be
| \eta | \ll 1 . \label{eta-small}
\ee
However, given that we are interested into understanding the effects of rapidly varying backgrounds, further derivatives of $\eta$ could be large, and the following hierarchy may be satisfied:
\be
| \eta |  \ll  |\tau \eta ' |  \ll  |\tau^2 \eta '' | . \label{hierarchy-1}
\ee
On the other hand, we may also consider rapid variations of the sound speed $c_s$ admitting departures from the slowly varying value $c_0 = 1$:
\be
\theta \equiv 1 - c_s^2 \ll 1 , \qquad | \theta |  \ll  |\tau \theta ' |  \ll  |\tau^2 \theta '' | . \label{hierarchy-s}
\ee
The hierarchies (\ref{hierarchy-1}) and (\ref{hierarchy-s}), together with eqs.~(\ref{epsilon-small}) and (\ref{eta-small}), reflect what we mean by having a rapid varying background near a quasi-de Sitter state. 

The previous assumptions allow us to rewrite $z''/z$ and $a''/ a$ in the following way
\be \label{eq:z-a}
\frac{z''}{z} = \frac{2}{\tau^2} \left(1+\frac{1}{2}\delta_{S}(\tau)\right),  \qquad  \frac{a''}{a} = \frac{2}{\tau^2}\left(1+\frac{1}{2}\delta_{T}(\tau)\right),
\ee
where we have used $\tau \simeq -(aH)^{-1}(1+\epsilon)$, and introduced the quantities $\delta_{S}(\tau)$ and $\delta_{T} (\tau)$ to parametrize the rapid variations of the background:
\be
\label{eq:deltas-def}
 \delta_{S}(\tau) = 3\epsilon+\frac{1}{2}\eta -\frac{\tau}{2}\eta' -3s + \tau s' , \qquad  \delta_{T}(\tau) = 3\epsilon .
\ee
By plugging these expressions back into the actions of eqs.~\eqref{S_s} and~\eqref{S_t} and treating the rapidly varying parts as interaction terms, we may split the theory as:
\bea \label{Sint-s}
S_{S}^0 &=& \frac{1}{2}\int d\tau d^3k \left[(u')^2 + k^2u^2 +  \frac{2}{\tau^2} u^2\right], \qquad S_{S}^{\rm int}=\frac{1}{2}\int d\tau d^3k \left[ \frac{\delta_{S}(\tau)}{\tau^2}u^2 \right], \\ \label{Sint-t}
S_{T}^0 &=& \frac{1}{2}\int d\tau d^3k \left[(f')^2 + k^2 f^2 +  \frac{2}{\tau^2}  f^2\right], \qquad S_{T}^{\rm int}=\frac{1}{2}\int d\tau d^3k \left[ \frac{\delta_{T}(\tau)}{\tau^2}f^2 \right].
\eea
Notice that eq.~(\ref{hierarchy-1}) implies a further hierarchy of the form
\be
| \delta | \ll | \tau \delta ' | \ll | \tau^2 \delta '' | ,  \label{hierarchy-2}
\ee
where $\delta$ stands for both $\delta_S$ and $\delta_T$. Given that a change in $e$-folds $dN$ is related to a change in conformal time by $dN = - d \tau / \tau$, the previous hierarchies simply tell us that $\delta_{S}$ and $\delta_{T}$ vary rapidly over an $e$-fold:
\be
|  \delta  | \ll \Big| \frac{ d \delta}{d N} \Big | \ll \Big| \frac{ d^2 \delta}{d N^2}  \Big| . \label{hierarchy-N}
\ee
As we shall see, these are the rapidly varying functions that source the appearance of features in the spectra.

\subsection{In-in formalism}

We may now use the standard \emph{in-in} formalism (see~\cite{Baumann:2009ds} for a review), which provides a way to compute the effects of the rapid time varying background on $n$-point correlation functions. To simplify the discussion, let us focus our attention on the scalar sector of the theory (\emph{i.e.} the $u$ fluctuations), and then come back to the case of tensor modes. Firstly, the complete solution $u (\textbf{k},\tau)$ can be written in terms of interaction picture fields $u_I (\textbf{k},\tau)$ as
\begin{eqnarray}
u(\textbf{k},\tau)&=&U_{}^{\dagger}(\tau)u_I(\textbf{k},\tau)U_{}(\tau),
\end{eqnarray}
where $U(\tau)$ is the propagator, given by
$$
U_{}(\tau)=\mathcal{T}\exp\left[-i\int^{\tau}_{-\infty_+}d\tau' H^{}_{I}(\tau')\right].
$$
Here $\mathcal{T}$ is the time ordering symbol, and $\infty_+=(1+i\epsilon)\infty$ is the usual prescription to choose the right vacuum in the infinite  past. In addition, $H^{}_{I}(\tau)$ is the interaction Hamiltonian, given by
\begin{eqnarray}
H^{}_I =  - \frac{\delta_{S}(\tau)}{\tau^2} \frac{1}{2}\int d^3k \,  u^2_{I} .
\end{eqnarray}
The interaction picture fields $u_I (\textbf{k},\tau)$ are given by free field solutions of the zeroth order action ({\emph i.e.} with $\delta_S = 0$), written in terms of creation and annihilation operators $a_{\k}^{\dag}$ and $a_{\k}$ as:
\be
u_I (\k, \tau) \equiv a_\k u_k(\tau)  + a_{-\k}^{\dag} u^{*}_k(\tau).
\ee
The creation and annihilation operators satisfy the standard commutation relation $\big[ a_{\k} , a_{\k'}^{\dag} \big] = (2 \pi)^3 \delta^{(3)} (\k - \k')$, whereas the mode functions $u_{k}(\tau)$ are given by mode solutions respecting Bunch-Davies initial conditions:
\be
u_{k} ( \tau ) = \frac{1}{\sqrt{2 k}} \left( 1 - \frac{i}{k \tau} \right) e^{-i k \tau} .
\ee 
Furthermore, the vacuum state $| 0 \rangle$ is defined to satisfy $a_{\k} | 0 \rangle = 0$. By expanding the propagator $U_{}(\tau)$, we may compute corrections to the two point function as
\be \label{2p-cor}
\langle u(\textbf{k},\tau)u(\textbf{k}',\tau) \rangle = \bra{0}u_I(\textbf{k},\tau)u_I(\textbf{k}',\tau) \ket{0}+i\int^\tau_{-\infty_+}  \!\!\!\!\!\! d\tau'\bra{0}[H_I(\tau'),u_I(\textbf{k},\tau)u_I(\textbf{k}',\tau) ]\ket{0} .
\ee
The power spectrum $\P_{\R}(k , \tau)$ of the primordial curvature perturbation $\R$ (evaluated at a given time $\tau$) is related to the two point function $\langle u(\textbf{k},\tau)u(\textbf{k}',\tau) \rangle$ as follows:
\be
\frac{1}{z^2} \langle u(\textbf{k},\tau)u(\textbf{k}',\tau) \rangle \equiv \frac{2 \pi^2}{k^3} \delta^{(3)} (\k - \k') \P_{\R}(k , \tau) .
\ee
We are interested in the power spectrum of super horizon modes at the end of inflation $\P_{\R}(k)$, which corresponds to the $\tau\to 0$ limit of $\P_{\R}(k , \tau) $. By taking into account the splitting of the theory into the zeroth order quasi-de Sitter part and the interaction part, we finally obtain
\be \label{ps_0}
\P_{\R}(k)=\P_{S}^0+\Delta \P_{S}(k), \qquad \P_S^0 (k)= \frac{H_0^2}{8\pi^2 \epsilon_0},
\ee
where $\P_{S}^0$ corresponds to the standard power spectrum for curvature perturbations in a quasi-de Sitter space-time, and $\Delta \P_{S}(k)$ contains the deviations from scale invariance induced by the rapidly varying background\footnote{Notice that in eq.~(\ref{eq:d_S}) time derivatives may be interchanged by factors of $- 2 i k$. Therefore, the appearance of four derivatives in $\theta$ might be deceiving, as the original expression~\cite{Palma:2014hra} leading to eq.~(\ref{eq:d_S}) had no time derivatives acting on $\theta$. Having time derivatives acting on both $\theta$ and $\delta_H$ in eq.~(\ref{eq:d_S}) allows one to have a single function of time being Fourier-transformed at the right hand side of the equation.}~\cite{Palma:2014hra}
\be \label{eq:d_S}
 \Delta_S(k) \equiv  \frac{\Delta \P_{S}}{\P^{0}_{S}} =\frac{i}{4 k^3 }\int_{-\infty}^\infty \! d\tau \left[\frac{\theta''''}{8} + \frac{\delta_{H}''}{2\tau^2}-\frac{\delta_H}{\tau^4} \right] e^{ 2 i k \tau },
\ee
where $\theta$ is defined in (\ref{hierarchy-s}) and $\delta_H$ is given by
\be \label{eq:d_H-def}
 \delta_H(\tau) = 3\epsilon+\frac{1}{2}\eta -\frac{\tau}{2}\eta'.
\ee 
Notice that the integration in eq.~(\ref{eq:d_S}) is performed over the whole real line $(- \infty , + \infty)$, which from now on will be omitted. To derive eq.~(\ref{eq:d_S}) we did the following trick~\cite{Achucarro:2012fd}: We extended the $\tau$-integration domain from $(-\infty, 0)$ to $(- \infty , + \infty)$ by imposing that both $\theta$ and $\delta_H$ are antisymmetric functions with respect to the interchange $\tau \to - \tau$.

We may now repeat all of the previous steps to compute the way that features appear in the tensor power spectrum. We find
$$
\P_{T}(k)=\P_{T}^0+\Delta \P_{T}(k), \qquad \P_T^0 (k)= \frac{H_0^2}{2\pi^2},
$$
where $\Delta \P_{T}(k)$ is given by
\be \label{eq:d_T}
\Delta_T(k) \equiv \frac{\Delta \P_{T}}{\P^{0}_{T}} = \frac{i}{4 k^3}\int  d\tau  \left[\frac{\delta_T''}{2\tau^2} -\frac{\delta_T}{\tau^4} \right] e^{2 i k\tau}.
\ee
Equations~(\ref{eq:d_S}) and~(\ref{eq:d_T}) are the basic equations that we will exploit to obtain the desired correlation between the two sectors of the theory. Before deducing such a relation, let us notice that the hierarchy of eq.~(\ref{hierarchy-2}) necessarily implies a hierarchy in Fourier space affecting the spectra, that reads
\be
| \Delta(k) | \ll \Big| \frac{d  \Delta(k)}{d \ln k}  \Big| \ll \Big| \frac{d^2 \Delta(k)}{d \ln k^2}  \Big| , \label{eq:hierarchy-Delta-k}
\ee
where $\Delta(k)$ stands for both $\Delta_S(k)$ and $\Delta_T(k)$.

\subsection{Features from varying Hubble parameters}

In this subsection we consider the case where $c_s = 1$ for all times, so that $\delta_S=\delta_H$, and any observable feature is the outcome of sudden variations of $H(t)$. Firstly, because of the hierarchy~(\ref{hierarchy-2}) satisfied by $\delta_T$, eq.~\eqref{eq:d_T} may be simplified as:
\be \label{eq:d_T2}
 \Delta_T(k) = \frac{i}{8 k^3 }\int \! d\tau \frac{\delta_T '' }{\tau^2 } e^{2ik\tau}.
\ee
Furthermore, because of eq.~(\ref{eq:deltas-def}), we see that eq.~(\ref{eq:d_T2}) may be rewritten in terms of $\eta$ as:
\be \label{eq:d_T3}
 \Delta_T(k) = - \frac{3 i \epsilon_0 }{8 k^3 }\int \! d\tau \frac{\eta' }{\tau^3 } e^{2ik\tau} . 
\ee
This expression may now be Fourier inverted, leading to a formal expression for $\eta'$ in terms of $\Delta_T(k)$ as 
\be
\eta'   =   \frac{ 1 }{3 \epsilon_0 } \int \! dk \left[ \frac{d^3}{d \ln k^3} \Delta_T(k) \right]  e^{-2ik\tau} . \label{eta-prime-Delta-T}
\ee
Next, we may use the hierarchy of eq.~(\ref{hierarchy-2}) satisfied by $\delta_S$ to rewrite eq.~\eqref{eq:d_S} as
\be \label{eq:d_S2}
 \Delta_S(k)  = - \frac{i}{16 k^3 }\int \! d\tau  \frac{1}{\tau} \eta'''  e^{2ik\tau} , 
\ee
where we used the fact that $\delta_{H} \simeq -\tau \eta' / 2$. As a last step, we may insert the expression for $\eta'$ in eq.~(\ref{eta-prime-Delta-T}) back into eq.~(\ref{eq:d_S2}), to obtain the main result of this work:
\be \label{eq:cor}
\frac{d^2}{d \ln k^2} \Delta_T = 6 \epsilon_0  \Delta_S.
\ee
This equation offers the desired link between features in the tensor and scalar spectra. Notice from eq.~(\ref{eq:d_T3}) that even though we have assumed that $\epsilon \ll 1$, the piece $\Delta_T(k)$ could in principle be large. However, from eq.~(\ref{eq:cor}), we see that features in the tensor power spectrum are highly suppressed with respect to those in the scalar spectrum. This is not only due to the presence of $\epsilon_0$~\cite{Hu:2014hoa}, but also due to the double $\ln k$-derivative acting on $\Delta_T(k)$, on account of the hierarchy (\ref{eq:hierarchy-Delta-k}). 

In the next subsection we extend this result to the more general case in which rapid variations of the sound speed are also allowed. As we shall see, in this case too, tensor features remain generically suppressed.

\subsection{Including the effects of a varying sound speed}

In the EFT of inflation~\cite{Cheung:2007st,Weinberg:2008hq}, the quadratic part of the action may exhibit a non-trivial sound speed for the perturbations, which could also lead to the presence of features in the scalar power spectrum~\cite{Achucarro:2010jv,Achucarro:2010da}. In general the evolution of $c_s(t)$ is independent of the evolution of $H$. That means that if features are generated by the simultaneous rapid variation of both $c_s$ and $H$, then the scalar and tensor power spectra would exhibit uncorrelated oscillatory features. This is because $\P_S$ would have features sourced by both $c_s$ and $H$ while $\P_T$ would have features sourced by $H$ alone. We would then have a relation of the form
\be
\Delta_S = \frac{1}{6 \epsilon_0} \frac{d^2}{d \ln k^2} \Delta_T  +  \Delta_{c} ,
\ee
where $\Delta_{c}$ represents the features sourced by variations of the sounds speed $c_s$.  

There are however intuitive reasons to expect that, at least in certain classes of models, variations of $c_s$ and $H$ happen in synchrony. An example of such a situation is the case where the inflationary valley admits turns, which is typical in multifield inflation~\cite{Achucarro:2010jv}. In these scenarios, as the inflaton traverses a curve in the field space, there are instant deviations from slow-roll produced by ``centrifugal" effects. Furthermore, the existence of such turns is responsible for a non-trivial sound speed~\cite{Tolley:2009fg}. The two quantities should thus be related since they stem from the same source. Another situation where $c_s$ and $H$ vary simultaneously is in $P(X,\varphi)$ models, where the kinetic term of the inflaton has a non-trivial structure. In these cases a reduction of the rapidity of the vacuum expectation value of the inflaton would inevitably induce a change in both $c_s$ and $H$.

To capture the aforementioned situations, in~\cite{Mooij:2015cxa}, a one parameter relation between the Hubble slow-roll parameter $\eta$ and the sound speed was proposed. This had the form
\be \label{alpha}
\eta = \eta_0 - \frac{\alpha}{2} \tau \theta',
\ee
with $\alpha \in \mathbb{R}$ and $\theta=1-c_s^2$. It was also shown to hold within several classes of models including $P(X,\varphi)$ and multifield models, with $\alpha$ admitting specific values for each case. 

Using this fact, one may now relate $\theta$ to $\eta$ in eq.~\eqref{eq:d_S} and follow the exact same steps to obtain a generic relation between the scalar and tensor power spectra in the case where both the sound speed and the Hubble radius experience sudden variations:
\be \label{eq:ode-cs}
\frac{d^2}{d \ln k^2} \Delta_T  = 6 \epsilon_0 \frac{\alpha}{1+\alpha} \Delta_S, \quad \alpha \neq -1,
\ee
and for the special case of $\alpha \simeq -1$:
\be \label{eq:ode-cs-2}
\frac{d}{d \ln k} \Delta_T = -\frac{6}{5} \epsilon_0 \Delta_S.
\ee
We see that in these set-up's too, deviations of the tensor power spectrum from scale invariance are suppressed by the slow-roll parameter $\epsilon$ as well as a double and a single momentum integral which smoothes out any acute variation of the scalar spectrum.

Before discussing quantitative features of these results, let us stress once more that the simple forms of eqs.~\eqref{eq:cor}, \eqref{eq:ode-cs} and \eqref{eq:ode-cs-2} are leading order expressions based on the assumption that any observable feature satisfy the following: \textit{i}) it is sharp, in the sense that any departure from scale invariance should take place within few e-folds, and \textit{ii}) it doesn't disrupt inflation, that is, $\epsilon$ remains small through out the whole dynamics.

\section{A quantitative discussion}  \label{sec:examples}

We now discuss the results of the previous section in two interesting situations. First, we consider the case in which resonant features are present throughout the whole spectra, and second, the case of the low $\ell$ power deficit observed in the scalar power spectrum. For this discussion, it will be useful to write concrete expressions relating features in the spectra and the rapidly varying contributions to the slow-roll parameters $\Delta \epsilon$ and $\Delta \eta$.  By Fourier inverting eq.~\eqref{eq:d_S2} for the general case where the sound speed also contributes to features, these are found to be given by~\cite{Palma:2014hra}
\be \label{eq:eta-e-inv}
\Delta \eta(\tau)= \frac{i}{\pi} \frac{\alpha}{1+\alpha} \int \! dk \left[ \frac{d}{dk} \frac{\Delta \P_S}{\P^0_{S}} \right] e^{2 i k\tau}  \;,\quad \Delta \epsilon(\tau)= \frac{i \epsilon_0}{ \pi } \frac{\alpha}{1+\alpha} \int \! dk \left[ \frac{1}{ k} \frac{\Delta \mathcal{P}_S}{\mathcal{P}^0_S} \right]  e^{2 i k\tau} ,
\ee
with $\Delta \epsilon$ following from the relation $\Delta \eta=-\tau \Delta \epsilon'/\epsilon_0$.
Note that the coefficient $\frac{\alpha}{1+\alpha} $ in eq.~\eqref{eq:eta-e-inv} is an ${\cal O}(1)$ number for any $\alpha$ so it's specific value has no impact on the results. We thus set it to one in what follows and work with eq.~\eqref{eq:cor}. The only case where it plays a role is when $\alpha \simeq-1$, in which the next to leading time derivative dominates in the RHS of eq.~\eqref{eq:d_S} leading to the following expressions:
\be \label{eq:eta-e-inv-a-1}
\Delta \eta(\tau)=- \frac{i}{5\pi}\int \! dk \, k \left[ \frac{d^2}{dk^2} \frac{\Delta \P_S}{\P^0_{S}} \right] e^{2 i k\tau} \;,\qquad 
\Delta \epsilon(\tau)= - \frac{i \epsilon_0}{5\pi} \int \! dk \left[ \frac{d}{ dk} \frac{\Delta \mathcal{P}_S}{\mathcal{P}^0_S}  \right] e^{2 i k\tau} .
\ee

\subsection{Resonant features}

This type of scale dependence is relevant in models of inflation where the potential is periodic or semi-periodic, such as axion monodromy inflation \cite{Flauger:2009ab}, or models like Natural Inflation \cite{Freese:1990rb}. Inflationary scenarios involving axions usually require super-Planckian field range, and hence, they are good candidates for the production of primordial gravitational waves \cite{Lyth:1996im,Easther:2006qu}. 

To acquire an idea of the possible impact of resonant features on the tensor power spectrum, we model the resonant part of the scalar power spectrum as
\be \label{eq:resonant}
\Delta_S(k) = A \cos \left( \Omega \log( k/k_*) + \phi \right),
\ee
where $A$ parametrizes the amplitude of the feature, while $\Omega$ and $\phi$ denote the frequency and the phase of the oscillation, respectively. To be concrete, we will consider the following values $A=0.028$, $\Omega=30$ and $\phi/2\pi=0.634$, which were found to constitute the best fit in the analysis of resonant features by Planck~\cite{Ade:2015lrj}. In addition, we set $k_*=0.05$ [Mpc]$^{-1}$ as a reference scale.

\subsubsection{Case for $\alpha \neq -1$}

\begin{figure}[H]
\begin{subfigure}{.5\textwidth}
  \centering
  \includegraphics[width=.8\linewidth]{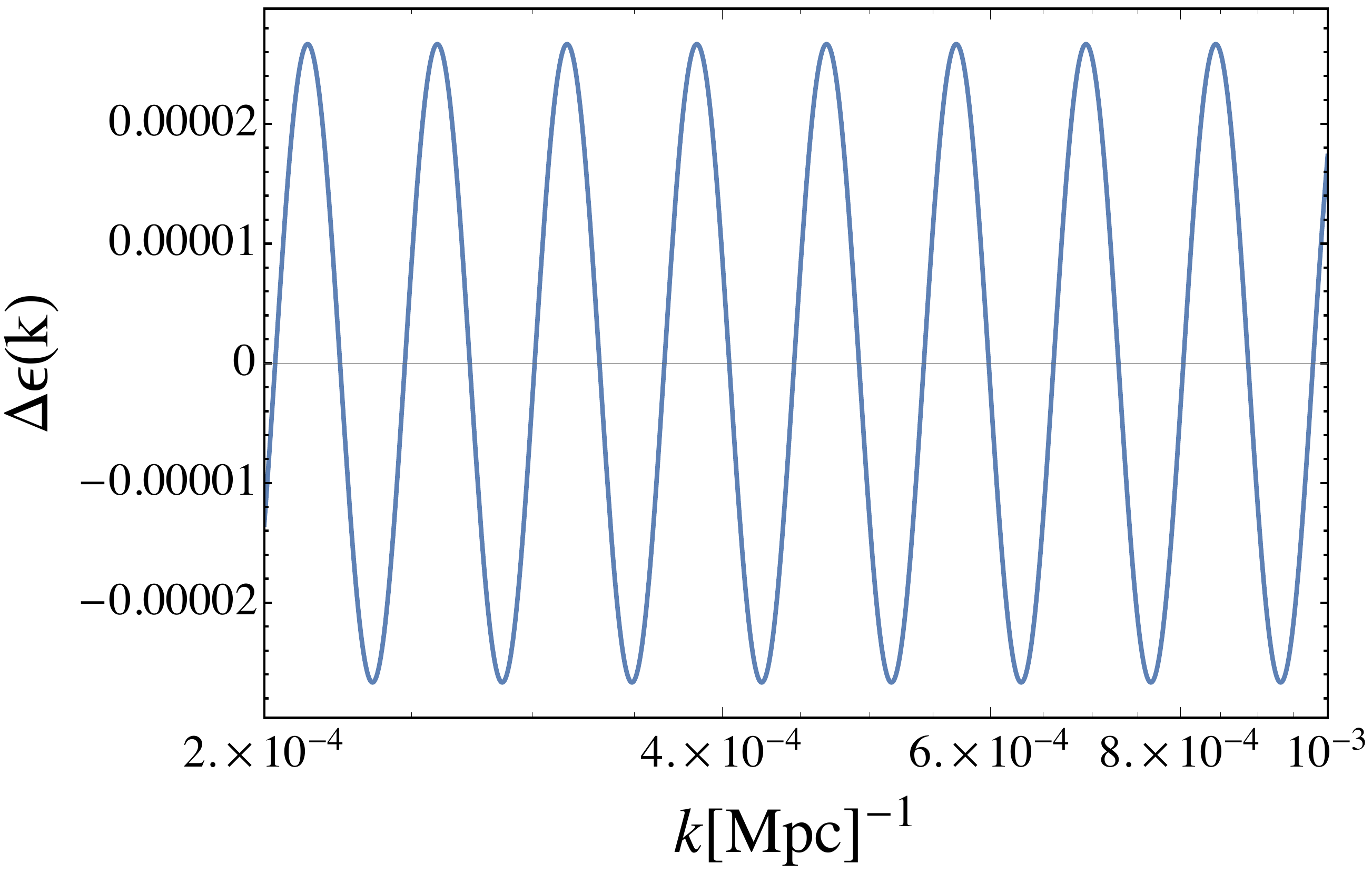}
  \label{fig:sfig2}
\end{subfigure}
\begin{subfigure}{.5\textwidth}
  \centering
  \includegraphics[width=.8\linewidth]{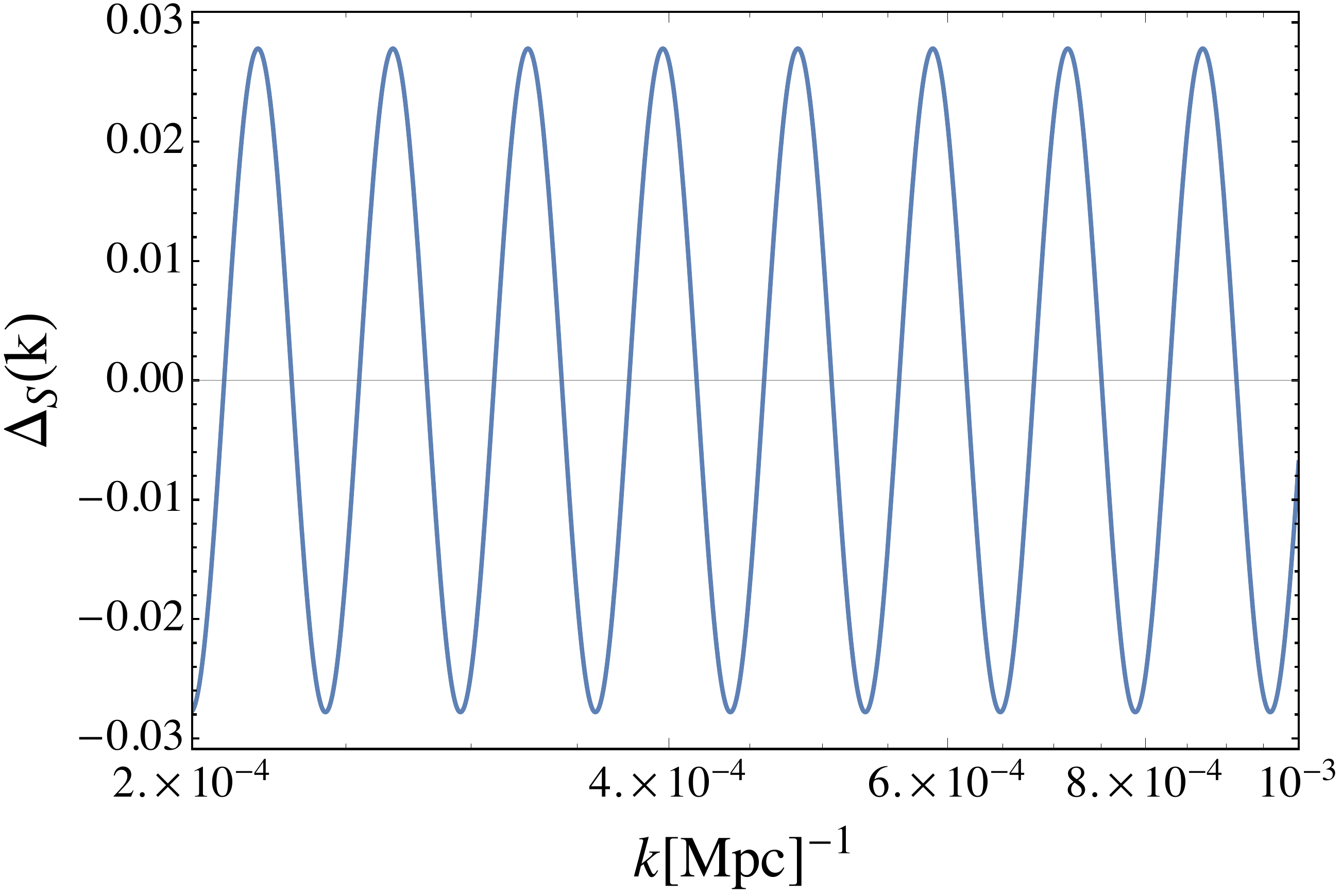}
  \label{fig:sfig3}
\end{subfigure}
\begin{subfigure}{.5\textwidth}
  \centering
  \includegraphics[width=.8\linewidth]{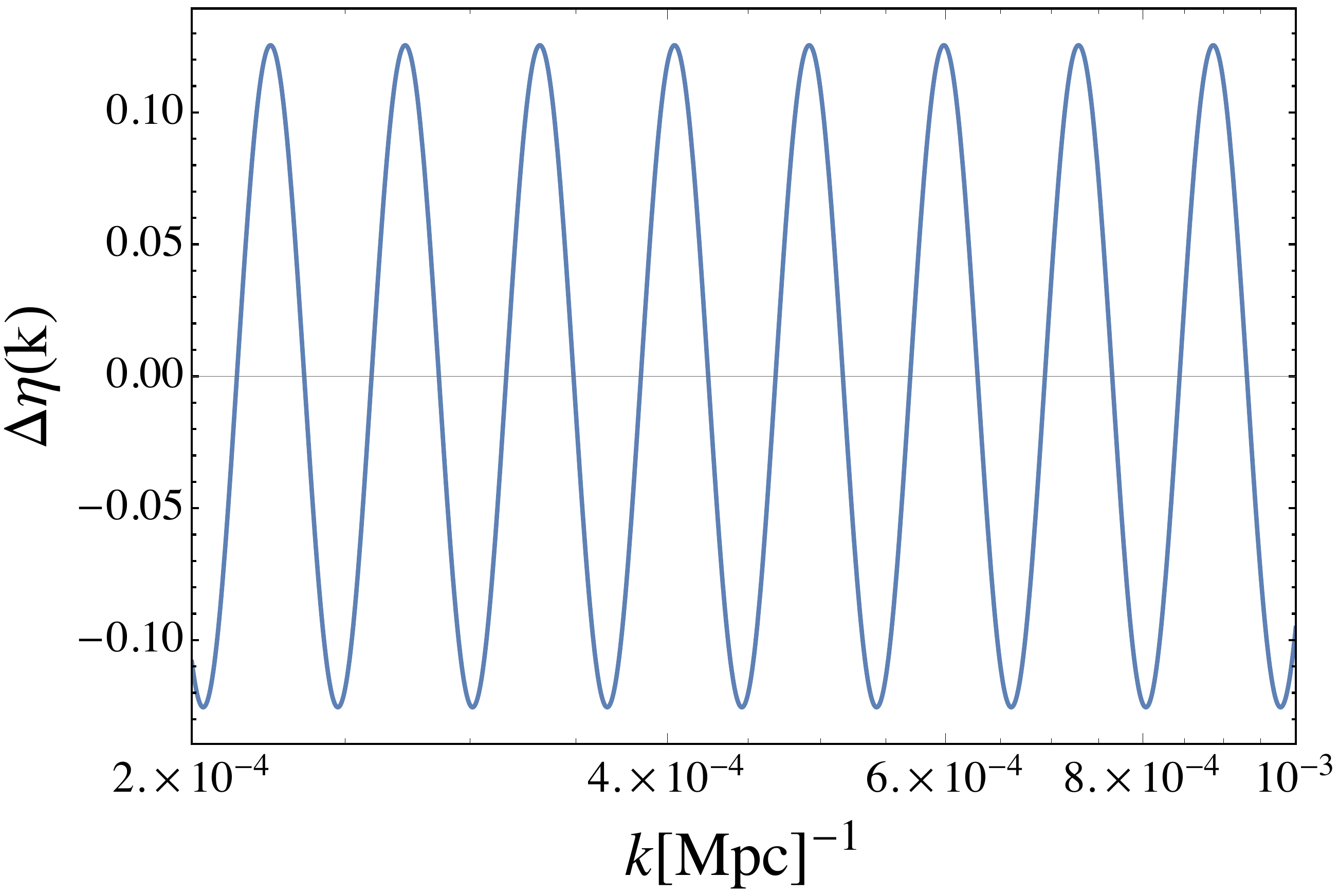}
  \label{fig:sfig2}
\end{subfigure}
\begin{subfigure}{.5\textwidth}
  \centering
  \includegraphics[width=.8\linewidth]{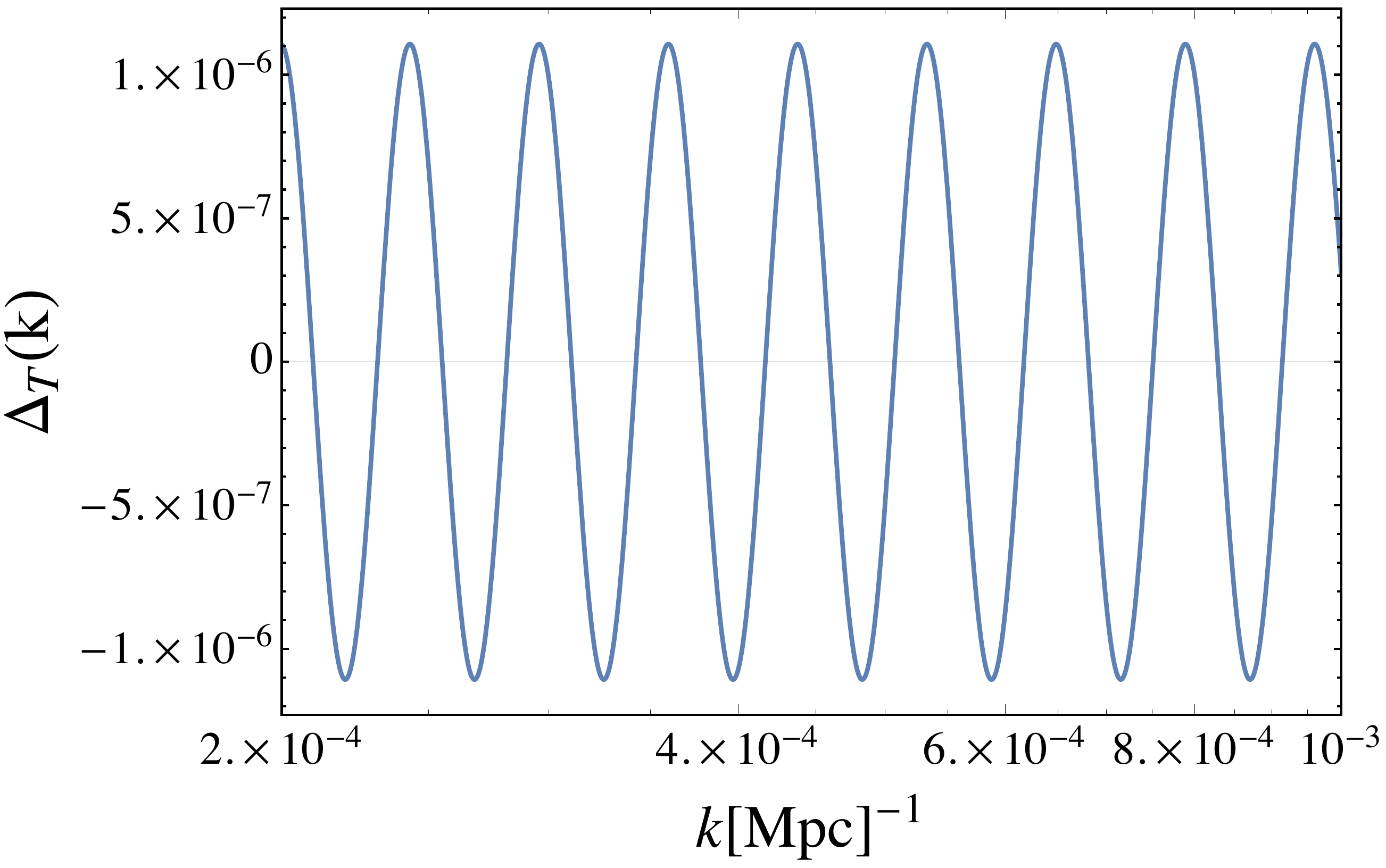}
  \label{fig:sfig1}
\end{subfigure}
\caption{Plot of the first two slow-roll parameters $\Delta \eta $ and $\Delta \epsilon$ (left panels) using eq.~\eqref{eq:eta-e-inv} and $\frac{ \Delta P_{S}}{P_{S}^0}(k)$, $\frac{ \Delta P_{T}}{P_{T}^0}(k)$ (right panels) related by eq.~\eqref{eq:cor}, in the case of the resonant feature \eqref{eq:resonant}. We have used $A=0.028$, $\Omega=30$, $\phi/2\pi=0.634$, $k_*=0.05$[Mpc]$^{-1}$ and $\epsilon_{0} = 0.0068$.}
\label{fig:res}
\end{figure}

Using the parametrization \eqref{eq:resonant} as a input, we numerically obtain the shape of the tensor spectrum feature via eq.~\eqref{eq:cor}, while the slow-roll parameters are reconstructed from eq.~\eqref{eq:eta-e-inv}. The results are shown in the plots of figure~\ref{fig:res}. There we see that features in the tensor power spectrum are present, albeit with an amplitude of $\Delta_T \sim 10^{-6}$ making them observationally irrelevant. This is a complementary argument in support of the claim that tensor features stemming from axionic potentials should be suppressed due to the smallness of the decay constant of the axion~\cite{Obata:2016xcr}.

\subsubsection{Case for $\alpha\simeq-1$}

\begin{figure}[H]
\begin{subfigure}{.5\textwidth}
  \centering
  \includegraphics[width=.8\linewidth]{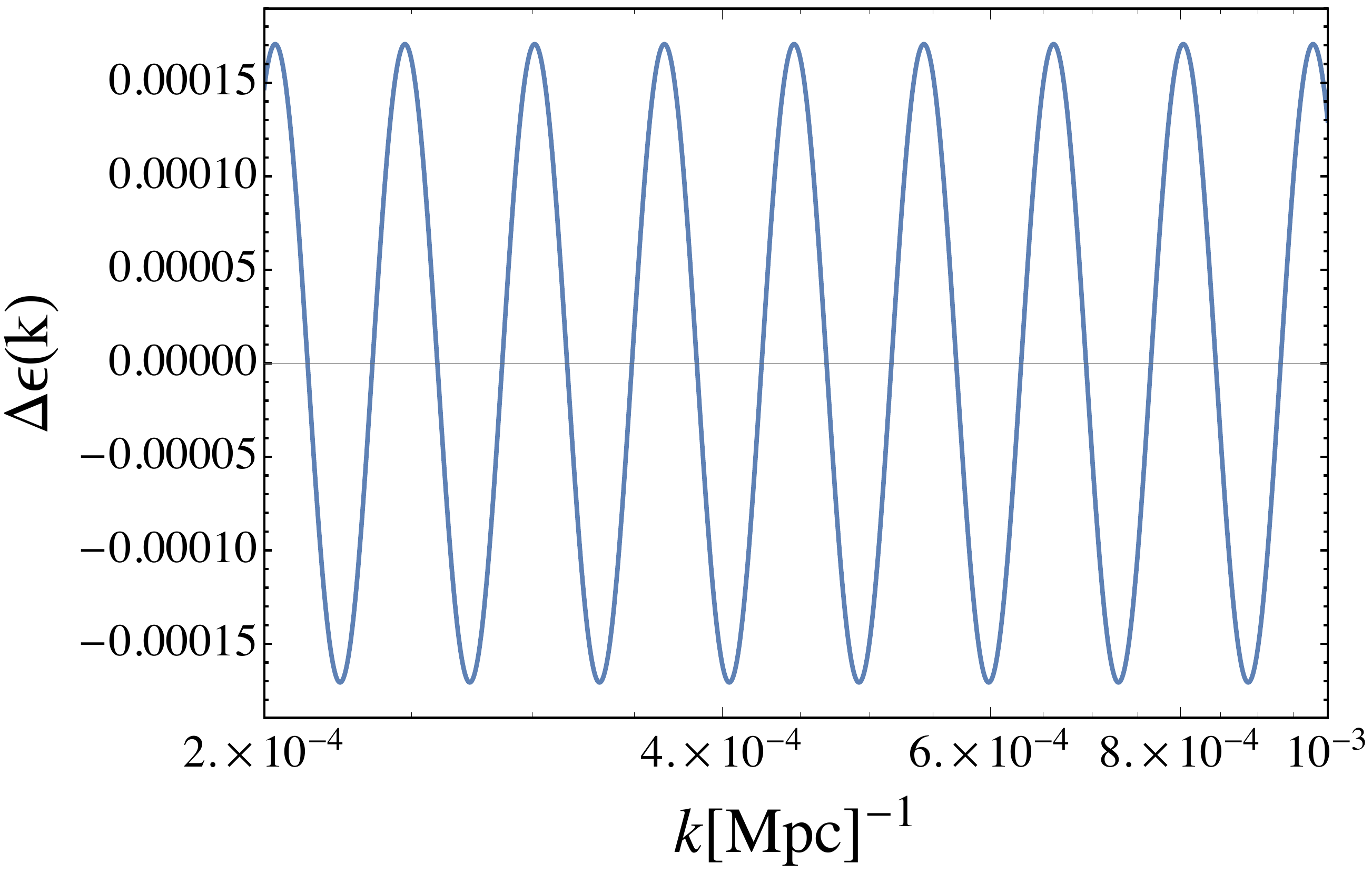}
  \label{fig:sfig2}
\end{subfigure}
\begin{subfigure}{.5\textwidth}
  \centering
  \includegraphics[width=.8\linewidth]{S-res.pdf}
  \label{fig:sfig3}
\end{subfigure}
\begin{subfigure}{.5\textwidth}
  \centering
  \includegraphics[width=.8\linewidth]{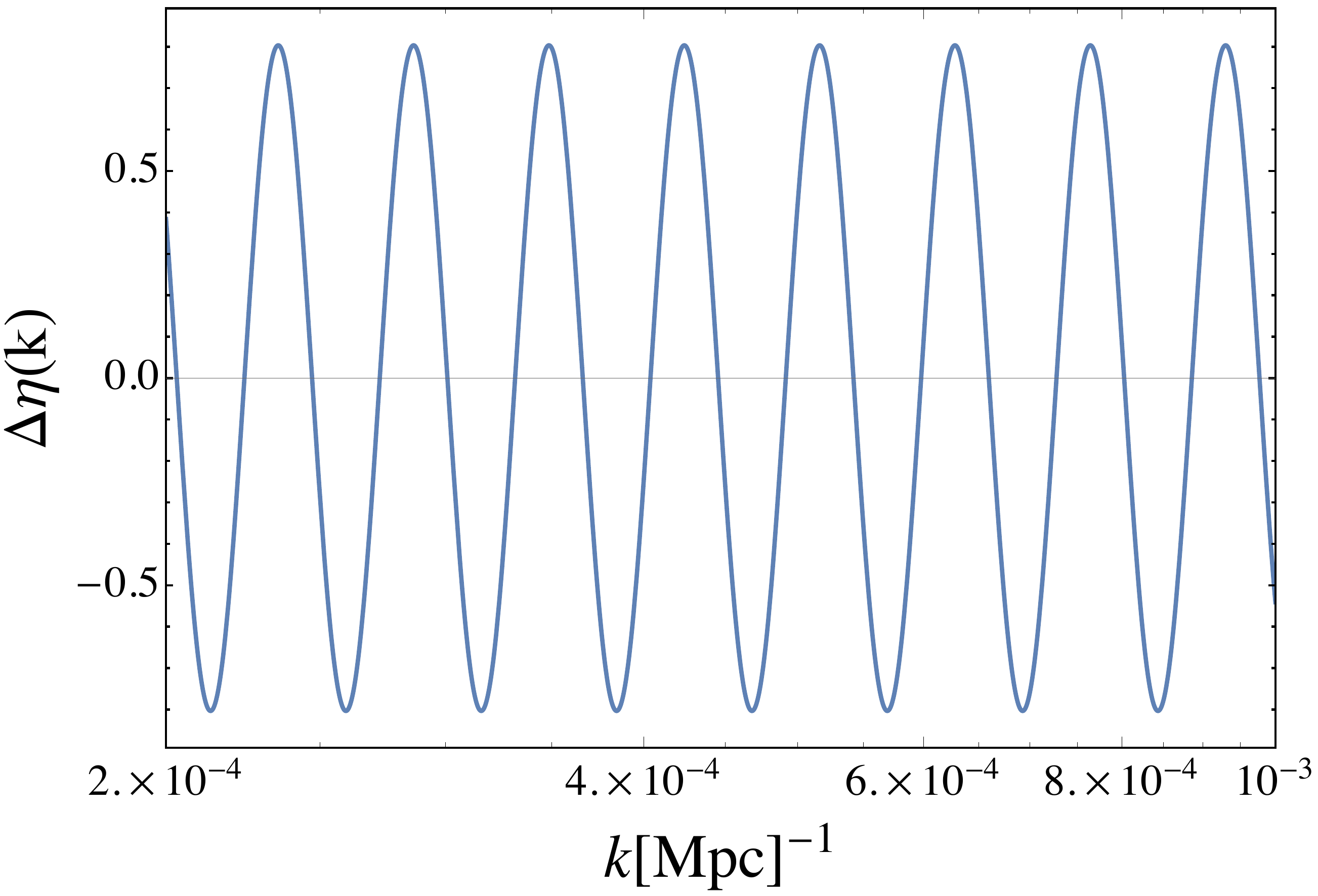}
  \label{fig:sfig2}
\end{subfigure}
\begin{subfigure}{.5\textwidth}
  \centering
  \includegraphics[width=.8\linewidth]{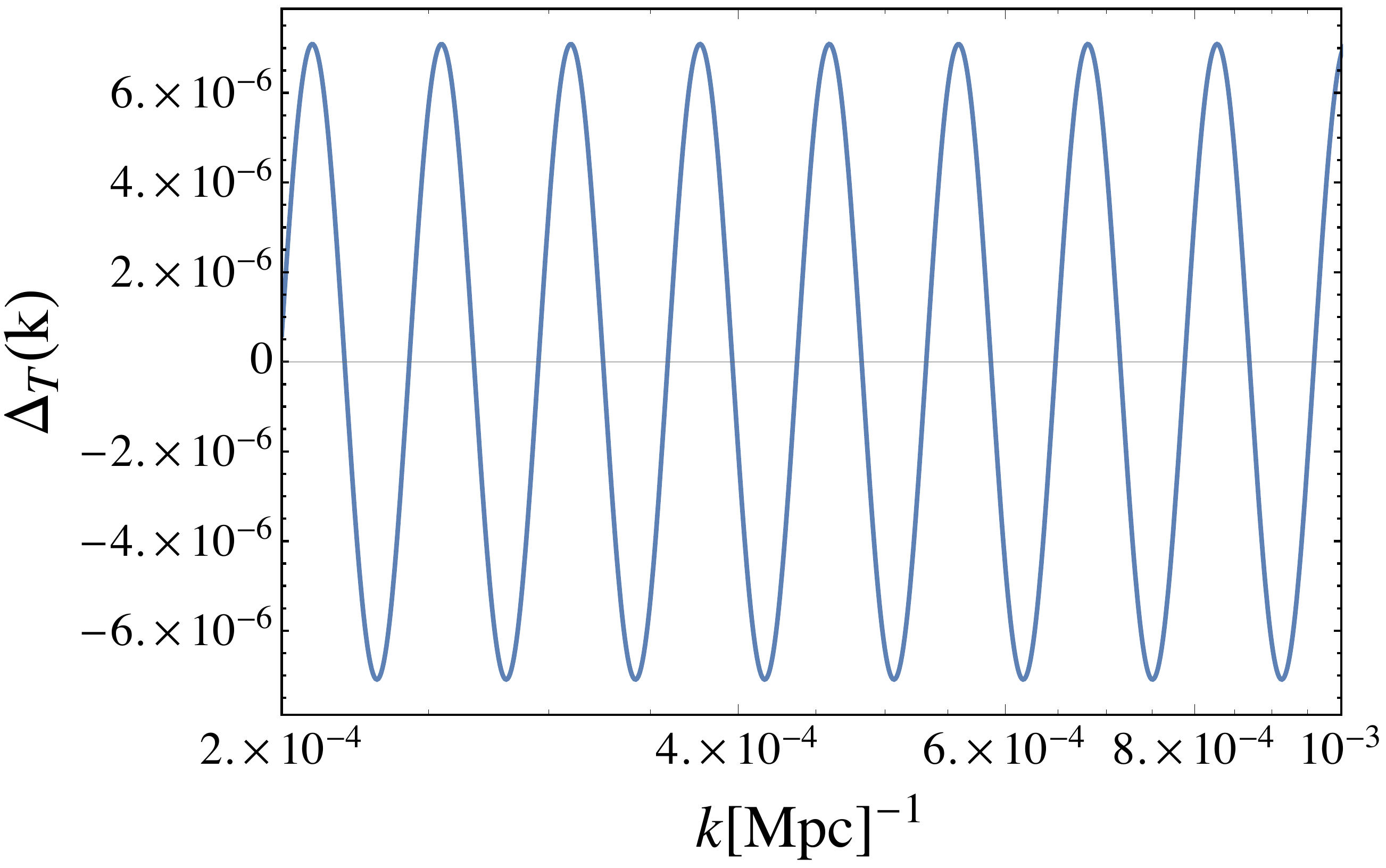}
  \label{fig:sfig1}
\end{subfigure}
\caption{Plot of the first two slow-roll parameters $\Delta \eta $ and $\Delta \epsilon$ (left panels) using eq.~\eqref{eq:eta-e-inv-a-1} and $\frac{ \Delta P_{S}}{P_{S}^0}(k)$, $\frac{ \Delta P_{T}}{P_{T}^0}(k)$ (right panels) related by eq.~\eqref{eq:ode-cs-2}, in the case of the resonant feature \eqref{eq:resonant}. We have used $A=0.028$, $\Omega=30$, $\phi/2\pi=0.634$, $k_*=0.05$[Mpc]$^{-1}$ and $\epsilon_{0} = 0.0068$.}
\label{fig:res-a}
\end{figure}

Next, we consider the special case of $\alpha\simeq-1$ for the resonance features. We numerically solve eqs.~\eqref{eq:ode-cs-2} and \eqref{eq:eta-e-inv-a-1}  and plot the results in figure~\ref{fig:res-a}. As can be seen, even though there is an order of magnitude enhancement with respect to the general case, the amplitude of the deviation from a scale invariant spectrum still remains extremely small. Furthermore, in this case $\eta$ can reach values up to $\eta \sim 0.8$. This does not invalidate the hierarchy \eqref{hierarchy-1}, as to go from eq.~(\ref{zppz}) to eq.~(\ref{eq:z-a}) one really requires $\eta / 2$ to be much smaller than $1$.

\subsection{Predictions for the low $\ell$ tensor power spectrum}

The low $\ell$ multipole region is the main observational window into CMB polarization since it is not contaminated by lensing effects. In addition, it is where the low $\ell$ deficit takes place in the scalar power spectrum~\cite{Aghanim:2015xee,Bennett:1996ce,Hinshaw:2003ex,Spergel:2003cb,Peiris:2003ff,Ade:2013zuv,Ade:2015xua,Ade:2013kta}. We focus in the $\ell < 50$ region, roughly corresponding to $0.0002 \lesssim k \lesssim 0.004$ [Mpc]$^{-1}$, which is the band that CMB polarization observatories focus on.

In order to get a quantitative look into the tensor power spectrum we model the $\ell\sim 20$ dip in the angular power spectrum as a sharp Gaussian:
\begin{equation} \label{gauss}
\Delta_S(k)=-A e^{-\lambda  (\ln(k/k_*))^2},
\end{equation}
where $k_*$ determines the location of the feature. We set $A=0.15$, $\lambda=15$ and $k_*=0.002$ [Mpc]$^{-1}$, which are chosen to have a rough fit with the observed power deficit. In addition, we choose $\epsilon_{0} = 0.0068$~\cite{Ade:2015lrj}. 

\subsubsection{Case with $\alpha \neq -1$}
\begin{figure}[H]
\begin{subfigure}{.5\textwidth}
  \centering
  \includegraphics[width=.8\linewidth]{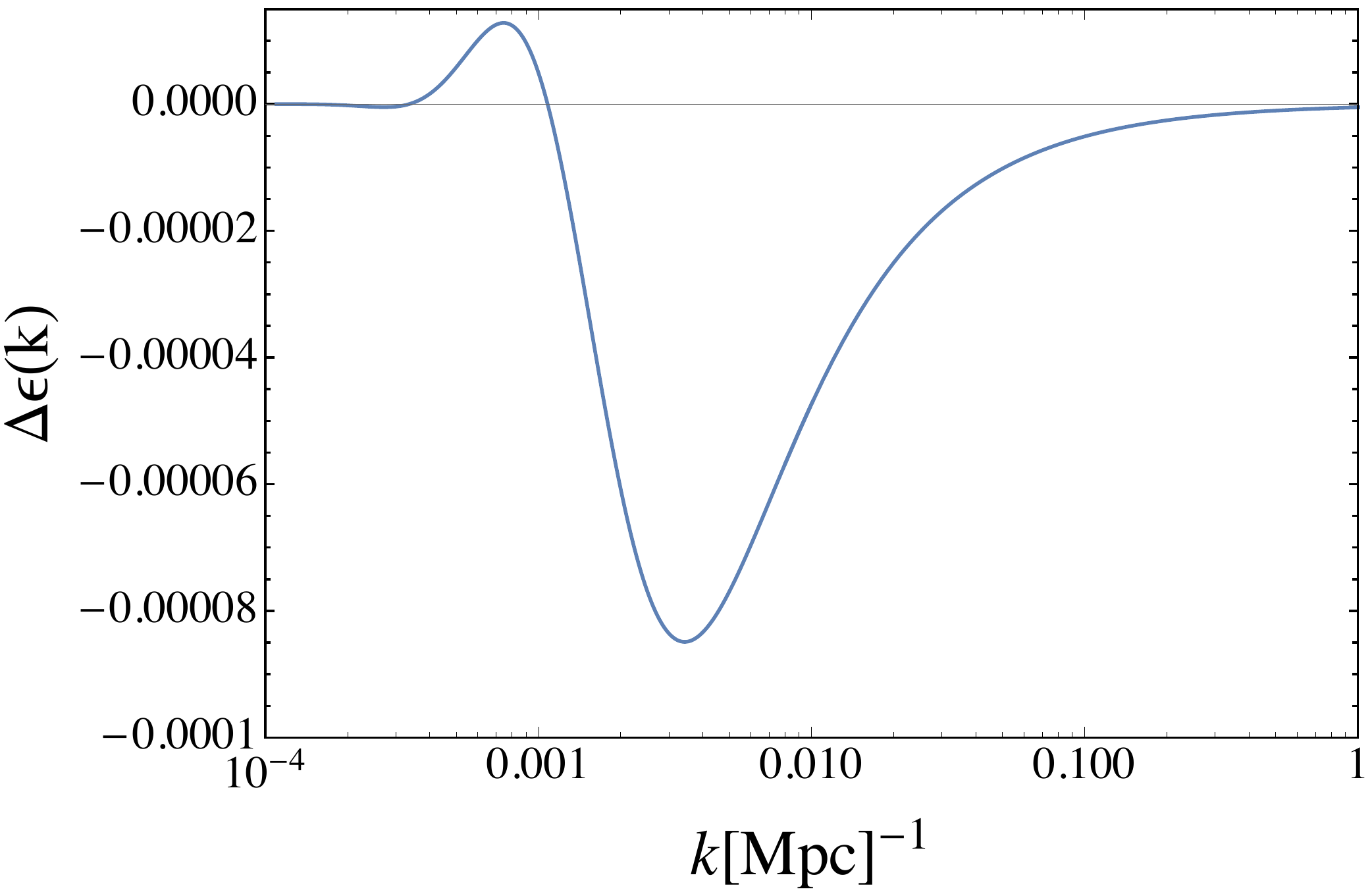}
  \label{fig:sfig2}
\end{subfigure}
\begin{subfigure}{.5\textwidth}
  \centering
  \includegraphics[width=.8\linewidth]{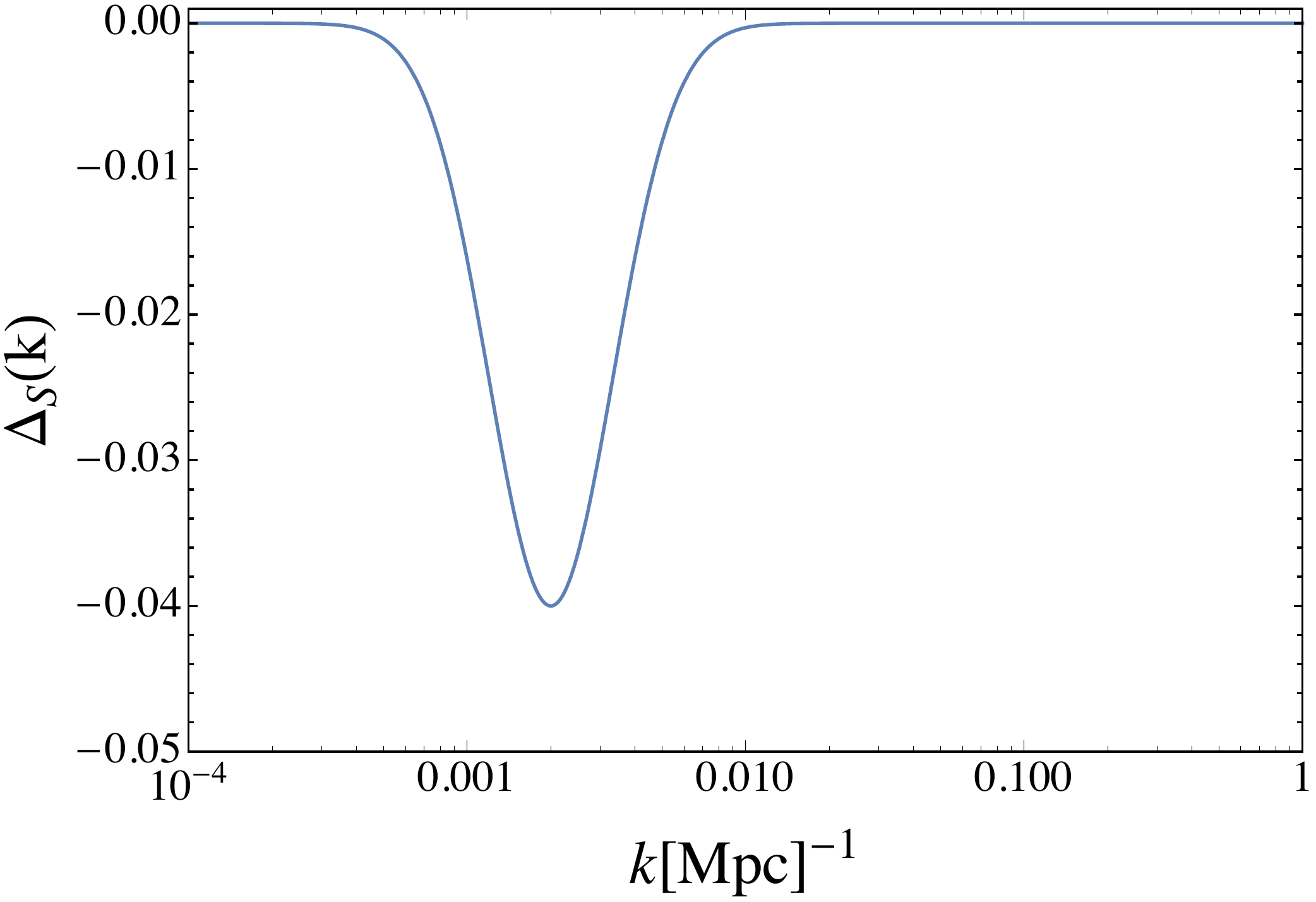}
  \label{fig:sfig3}
\end{subfigure}
\begin{subfigure}{.5\textwidth}
  \centering
  \includegraphics[width=.8\linewidth]{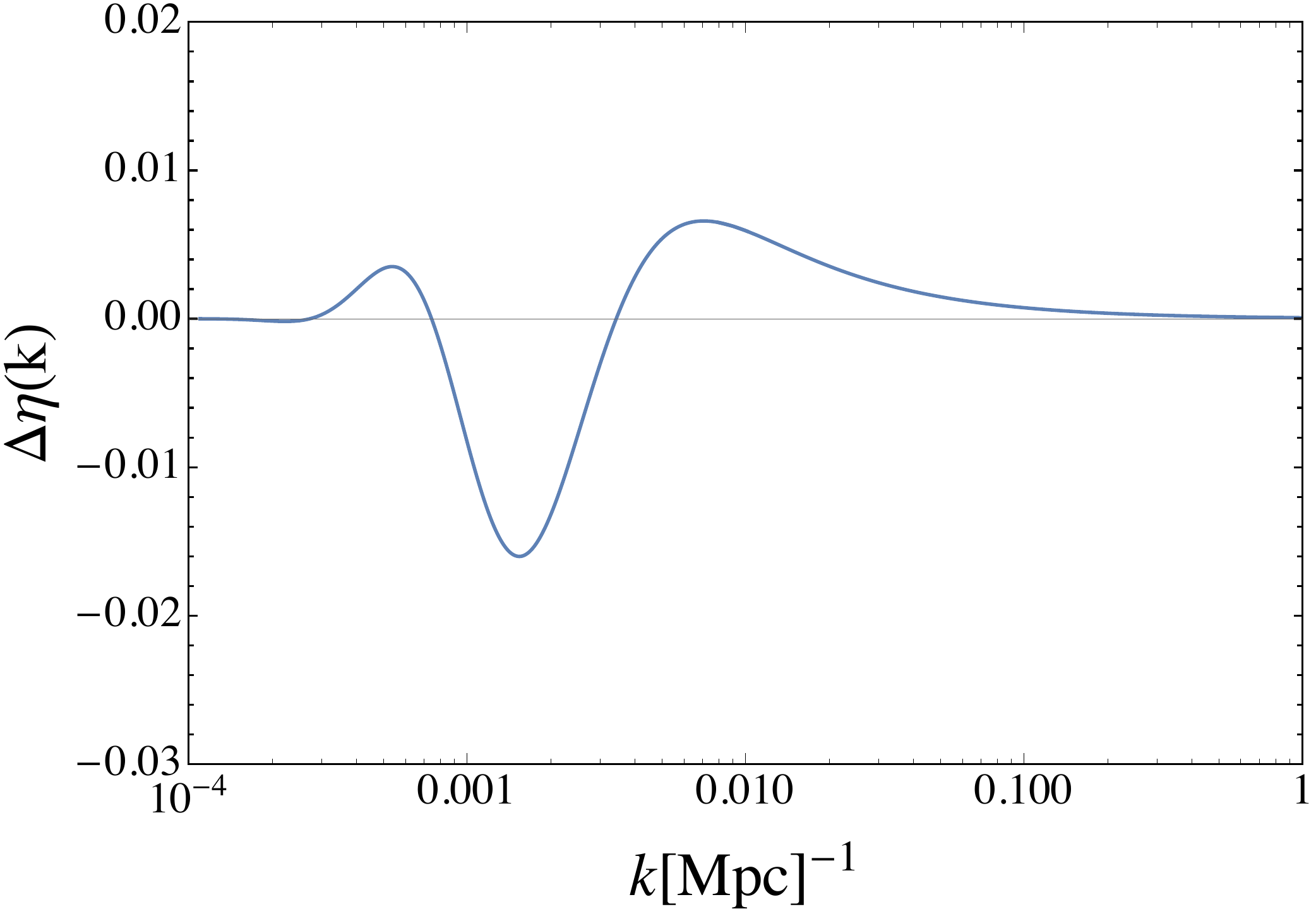}
  \label{fig:sfig2}
\end{subfigure}
\begin{subfigure}{.5\textwidth}
  \centering
  \includegraphics[width=.8\linewidth]{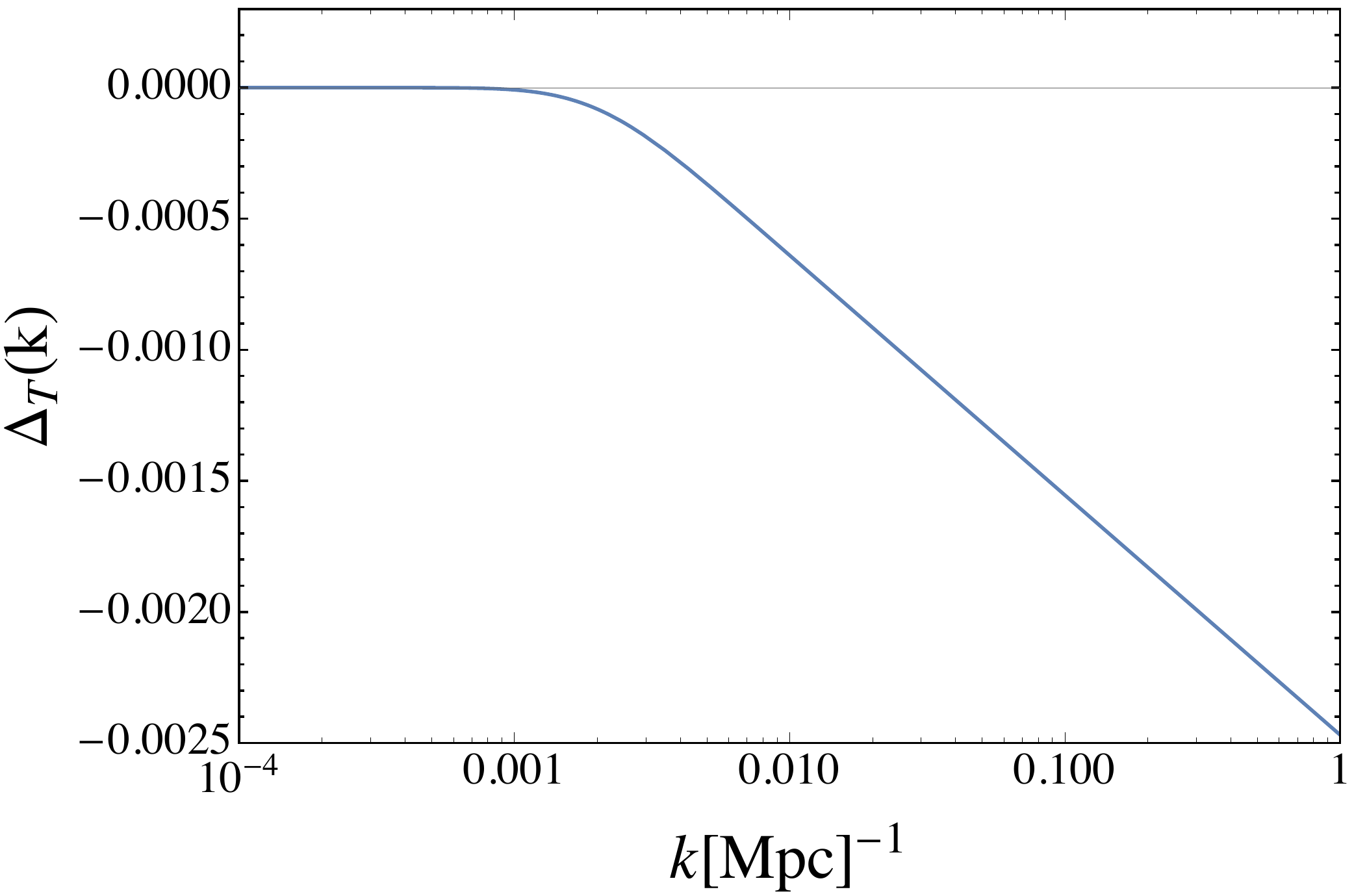}
  \label{fig:sfig1}
\end{subfigure}
\caption{Plot of the first two slow-roll parameters $\Delta \eta $ and $\Delta \epsilon$ (left panels) using eq.~\eqref{eq:eta-e-inv} and $\frac{ \Delta P_{S}}{P_{S}^0}(k)$, $\frac{ \Delta P_{T}}{P_{T}^0}(k)$ (right panels) related by eq.~\eqref{eq:cor}, in the case of the Gaussian feature \eqref{gauss}. We have used $A=-0.15$, $\lambda= 15$, $ k^{*}=0.002$[Mpc]$^{-1}$ and $\epsilon_{0} = 0.0068$.}
\label{fig:g}
\end{figure}

We solve eqs.~\eqref{eq:cor} and \eqref{eq:eta-e-inv} with the parametrization \eqref{gauss} as an input, with the results shown in the plots of figure~\ref{fig:g}. We see that for a realistic amplitude $A$ the tensor power spectrum exhibits a feature of amplitude $\Delta_T \sim 10^{-9}.$ 

\subsubsection{Case with $\alpha\simeq-1$}

\begin{figure}[H]
\begin{subfigure}{.5\textwidth}
  \centering
  \includegraphics[width=.8\linewidth]{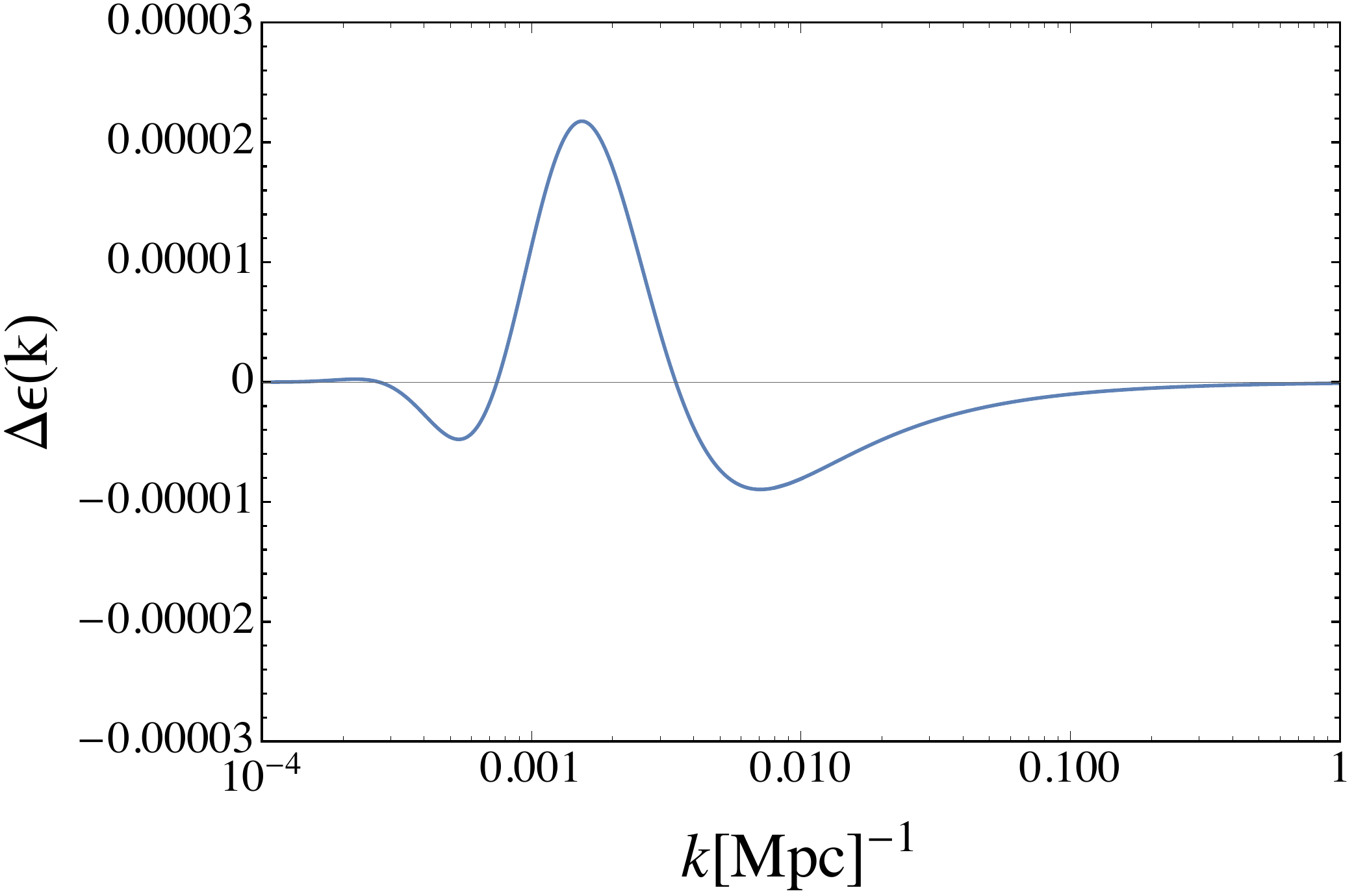}
  \label{fig:sfig2}
\end{subfigure}
\begin{subfigure}{.5\textwidth}
  \centering
  \includegraphics[width=.8\linewidth]{S-g.pdf}
  \label{fig:sfig3}
\end{subfigure}
\begin{subfigure}{.5\textwidth}
  \centering
  \includegraphics[width=.8\linewidth]{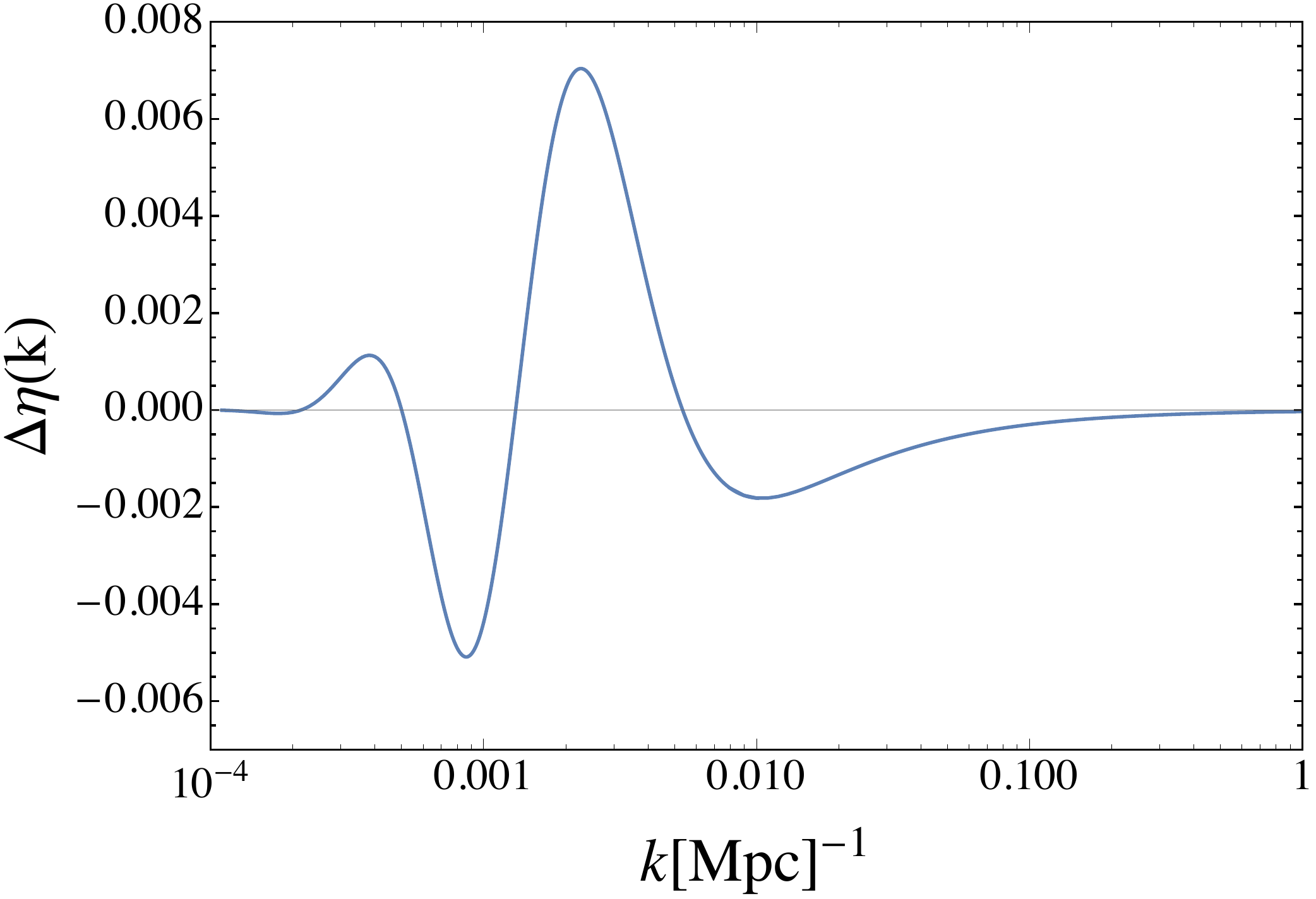}
  \label{fig:sfig2}
\end{subfigure}
\begin{subfigure}{.5\textwidth}
  \centering
  \includegraphics[width=.8\linewidth]{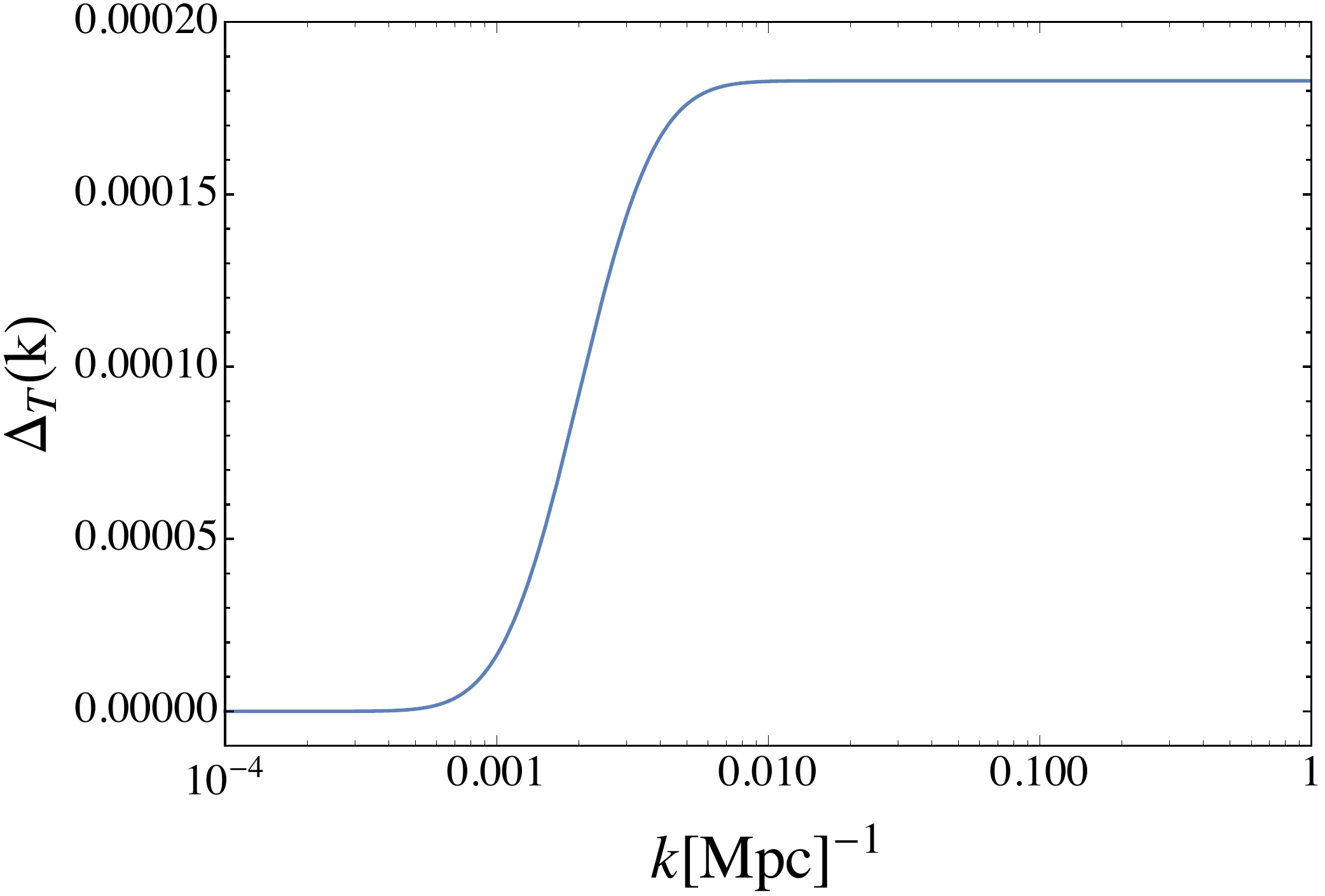}
  \label{fig:sfig1}
\end{subfigure}
\caption{Plot of the first two slow-roll parameters $\Delta \eta $ and $\Delta \epsilon$ (left panels) using eq.~\eqref{eq:eta-e-inv-a-1} and $\frac{ \Delta P_{S}}{P_{S}^0}(k)$, $\frac{ \Delta P_{T}}{P_{T}^0}(k)$ (right panels) related by eq.~\eqref{eq:ode-cs-2}, in the case of the Gaussian feature \eqref{gauss}. We have used $A=-0.15$, $\lambda= 15$, $ k^{*}=0.002$[Mpc]$^{-1}$ and $\epsilon_{0} = 0.0068$.}
\label{fig:g-a}
\end{figure}

In the special case of $\alpha\simeq-1$, we see that the tensor spectrum and the slow-roll parameters, now given by eqs.~\eqref{eq:ode-cs-2} and \eqref{eq:eta-e-inv-a-1} respectively, exhibit a feature which is enhanced by an order of magnitude compared to the previous case. However, as seen in figure~\ref{fig:g-a}, the amplitude still remains extremely small.

\section{Conclusions} \label{sec:conclusions}
We have studied the possible appearance of scale dependent features in the power spectrum of primordial tensor perturbations due to non-trivial inflationary dynamics in a model independent way. Our main result is eq.~\eqref{eq:cor} -- or eqs.~(\ref{eq:ode-cs}),~\eqref{eq:ode-cs-2} in the more general case of EFT's with a sound speed -- which consist of relations linking features in the tensor power spectrum to those appearing in the scalar power spectrum, allowing us to estimate the amplitude and shape of the former given the latter. In general, we find that the tensor spectrum is expected to be featureless: Indeed, eq.~\eqref{eq:cor} shows that any feature appearing in the tensor spectrum is generically suppressed with respect to those appearing in the scalar one for two reasons: firstly due to slow-roll~\cite{Hu:2014hoa}, and more importantly, due to the fact that features should in general be sharp enough in order to leave an imprint in the CMB.

One may wonder about other mechanisms producing features in the tensor sector of the theory. For instance, in principle, we could consider a Lagrangian describing the dynamics of tensor modes with a sound speed $c_t$ experiencing rapid variations producing features in the tensor spectrum. However, in~\cite{Creminelli:2014wna} it was shown that under a disformal transformation, models with a non-trivial tensor sound speed (and canonical scalar sector) map into models with a non-trivial scalar sound speed (and canonical tensor sector). Since the spectra are invariant under such a transformation, our formalism to relate features in the tensor spectrum to those appearing in the scalar spectrum would continue to be valid. Moreover, in the special case where only $c_t$ varies, the disformal transformation would lead to an equivalent system where both $c_s$ and $H$ vary, but in such a way that the scalar spectrum remains featureless~\cite{Cai:2015yza}. Given that we are interested in understanding the consequences of features in the scalar spectrum on the tensor one, this class of situations is out of our scope.

Current CMB observations show the existence of departures from scale invariance in the power spectrum of primordial curvature perturbations in the multipole range $\ell \sim 20$. If we interpret this behavior as the result of the dynamics of inflation, we are led to conclude that the tensor power spectrum will not show any consequential departure from scale invariance in this region. The importance of this conclusion may be appreciated more clearly by inverting the statement: If tensor modes are observed to have strong departures from scale invariance in the aforementioned multipole range, then we will have good reasons to suspect that the departures appearing in the scalar spectrum are not of primordial origin.

\subsection*{Acknowledgements}
This work is supported in part by Fondecyt project 1130777 (GAP, BP, WR) and Fondecyt 2016 Post-doctoral Grant 3160299 (SS). BP acknowledges support from the CONICYT postgraduate scholarship program. WR acknowledges support from the DFI postgraduate scholarship program.

\section*{}
%
%
%
%

%
%
%
%
%
%
%
%


\begin{thebibliography}{99}


\bibitem{Guth:1980zm} 
  A.~H.~Guth,
  ``The Inflationary Universe: A Possible Solution to the Horizon and Flatness Problems,''
  Phys.\ Rev.\ D {\bf 23}, 347 (1981).
  
\bibitem{Linde:1981mu} 
  A.~D.~Linde,
  ``A New Inflationary Universe Scenario: A Possible Solution of the Horizon, Flatness, Homogeneity, Isotropy and Primordial Monopole Problems,''
  Phys.\ Lett.\ B {\bf 108}, 389 (1982).
  
\bibitem{Starobinsky:1980te} 
  A.~A.~Starobinsky,
  ``A New Type of Isotropic Cosmological Models Without Singularity,''
  Phys.\ Lett.\ B {\bf 91}, 99 (1980).

\bibitem{Albrecht:1982wi} 
  A.~Albrecht and P.~J.~Steinhardt,
  ``Cosmology for Grand Unified Theories with Radiatively Induced Symmetry Breaking,''
  Phys.\ Rev.\ Lett.\  {\bf 48}, 1220 (1982).

\bibitem{Abazajian:2016yjj} 
  K.~N.~Abazajian {\it et al.} [CMB-S4 Collaboration],
  ``CMB-S4 Science Book, First Edition,''
  arXiv:1610.02743 [astro-ph.CO].

\bibitem{Harrington:2016jrz} 
  K.~Harrington {\it et al.},
  ``The Cosmology Large Angular Scale Surveyor,''
  Proc.\ SPIE Int.\ Soc.\ Opt.\ Eng.\  {\bf 9914}, 99141K (2016)
  [arXiv:1608.08234 [astro-ph.IM]].

\bibitem{Suzuki:2015zzg} 
  A.~Suzuki {\it et al.} [POLARBEAR Collaboration],
  ``The POLARBEAR-2 and the Simons Array Experiment,''
  J.\ Low.\ Temp.\ Phys.\  {\bf 184}, no. 3-4, 805 (2016)
  [arXiv:1512.07299 [astro-ph.IM]].

\bibitem{Ahmed:2014ixy} 
  Z.~Ahmed {\it et al.} [BICEP3 Collaboration],
  ``BICEP3: a 95GHz refracting telescope for degree-scale CMB polarization,''
  Proc.\ SPIE Int.\ Soc.\ Opt.\ Eng.\  {\bf 9153}, 91531N (2014)
  [arXiv:1407.5928 [astro-ph.IM]].
  
  

\bibitem{Aghanim:2015xee} 
  N.~Aghanim {\it et al.} [Planck Collaboration],
  ``Planck 2015 results. XI. CMB power spectra, likelihoods, and robustness of parameters,''
  Astron.\ Astrophys.\  {\bf 594}, A11 (2016)
  [arXiv:1507.02704 [astro-ph.CO]].
  
  
\bibitem{Hazra:2013nca} 
  D.~K.~Hazra, A.~Shafieloo and G.~F.~Smoot,
  ``Reconstruction of broad features in the primordial spectrum and inflaton potential from Planck,''
  JCAP {\bf 1312}, 035 (2013)
  [arXiv:1310.3038 [astro-ph.CO]].
  
\bibitem{Hunt:2013bha} 
  P.~Hunt and S.~Sarkar,
  ``Reconstruction of the primordial power spectrum of curvature perturbations using multiple data sets,''
  JCAP {\bf 1401}, 025 (2014)
  [arXiv:1308.2317 [astro-ph.CO]].
  
\bibitem{Hazra:2014jwa} 
  D.~K.~Hazra, A.~Shafieloo and T.~Souradeep,
  ``Primordial power spectrum from Planck,''
  JCAP {\bf 1411}, no. 11, 011 (2014)
  [arXiv:1406.4827 [astro-ph.CO]].

\bibitem{Hunt:2015iua} 
  P.~Hunt and S.~Sarkar,
  ``Search for features in the spectrum of primordial perturbations using Planck and other datasets,''
  JCAP {\bf 1512}, no. 12, 052 (2015)
  [arXiv:1510.03338 [astro-ph.CO]].
  
  

  


\bibitem{Bennett:1996ce} 
  C.~L.~Bennett {\it et al.},
  ``Four year COBE DMR cosmic microwave background observations: Maps and basic results,''
  Astrophys.\ J.\  {\bf 464}, L1 (1996)
  [astro-ph/9601067].
  
  
\bibitem{Hinshaw:2003ex} 
  G.~Hinshaw {\it et al.} [WMAP Collaboration],
  ``First year Wilkinson Microwave Anisotropy Probe (WMAP) observations: The Angular power spectrum,''
  Astrophys.\ J.\ Suppl.\  {\bf 148}, 135 (2003)
  [astro-ph/0302217].
  
  
\bibitem{Spergel:2003cb} 
  D.~N.~Spergel {\it et al.} [WMAP Collaboration],
  ``First year Wilkinson Microwave Anisotropy Probe (WMAP) observations: Determination of cosmological parameters,''
  Astrophys.\ J.\ Suppl.\  {\bf 148}, 175 (2003)
  [astro-ph/0302209].
  
  
\bibitem{Peiris:2003ff} 
  H.~V.~Peiris {\it et al.} [WMAP Collaboration],
  ``First year Wilkinson Microwave Anisotropy Probe (WMAP) observations: Implications for inflation,''
  Astrophys.\ J.\ Suppl.\  {\bf 148}, 213 (2003)
  [astro-ph/0302225].
  
  
\bibitem{Ade:2013zuv} 
  P.~A.~R.~Ade {\it et al.} [Planck Collaboration],
  ``Planck 2013 results. XVI. Cosmological parameters,''
  Astron.\ Astrophys.\  {\bf 571}, A16 (2014)
  [arXiv:1303.5076 [astro-ph.CO]].
  
\bibitem{Ade:2015xua} 
  P.~A.~R.~Ade {\it et al.} [Planck Collaboration],
  ``Planck 2015 results. XIII. Cosmological parameters,''
  Astron.\ Astrophys.\  {\bf 594}, A13 (2016)
  [arXiv:1502.01589 [astro-ph.CO]].
  
\bibitem{Ade:2013kta} 
  P.~A.~R.~Ade {\it et al.} [Planck Collaboration],
  ``Planck 2013 results. XV. CMB power spectra and likelihood,''
  Astron.\ Astrophys.\  {\bf 571}, A15 (2014)
  [arXiv:1303.5075 [astro-ph.CO]].



\bibitem{Benetti:2016tvm} 
  M.~Benetti and J.~S.~Alcaniz,
  ``Bayesian analysis of inflationary features in Planck and SDSS data,''
  Phys.\ Rev.\ D {\bf 94}, no. 2, 023526 (2016)
  [arXiv:1604.08156 [astro-ph.CO]].
  
  
\bibitem{Benetti:2013cja} 
  M.~Benetti,
  ``Updating constraints on inflationary features in the primordial power spectrum with the Planck data,''
  Phys.\ Rev.\ D {\bf 88}, 087302 (2013)
  [arXiv:1308.6406 [astro-ph.CO]].
  
\bibitem{Novaes:2015uza} 
  C.~P.~Novaes, M.~Benetti and A.~Bernui,
  ``Primordial Non-Gaussianities of inflationary step-like models,''
  arXiv:1507.01657 [astro-ph.CO].
  
  
  
  
  
\bibitem{Ashoorioon:2008qr} 
  A.~Ashoorioon, A.~Krause and K.~Turzynski,
  ``Energy Transfer in Multi Field Inflation and Cosmological Perturbations,''
  JCAP {\bf 0902}, 014 (2009)
  [arXiv:0810.4660 [hep-th]].
  
  
\bibitem{Gariazzo:2016blm} 
  S.~Gariazzo, O.~Mena, H.~Ramirez and L.~Boubekeur,
  ``Primordial power spectrum features in phenomenological descriptions of inflation,''
  arXiv:1606.00842 [astro-ph.CO].
  
  
\bibitem{Gariazzo:2015qea} 
  S.~Gariazzo, L.~Lopez-Honorez and O.~Mena,
  ``Primordial Power Spectrum features and $f_{NL}$ constraints,''
  Phys.\ Rev.\ D {\bf 92}, no. 6, 063510 (2015)
  [arXiv:1506.05251 [astro-ph.CO]].
  
  
\bibitem{Gao:2015aba} 
  X.~Gao and J.~O.~Gong,
  ``Towards general patterns of features in multi-field inflation,''
  JHEP {\bf 1508}, 115 (2015)
  [arXiv:1506.08894 [astro-ph.CO]].
  
\bibitem{Cai:2015xla} 
  Y.~F.~Cai, E.~G.~M.~Ferreira, B.~Hu and J.~Quintin,
  ``Searching for features of a string-inspired inflationary model with cosmological observations,''
  Phys.\ Rev.\ D {\bf 92}, no. 12, 121303 (2015)
  [arXiv:1507.05619 [astro-ph.CO]].
  
  
\bibitem{GallegoCadavid:2016wcz} 
  A.~Gallego Cadavid, A.~E.~Romano and S.~Gariazzo,
  ``CMB anomalies and the effects of local features of the inflaton potential,''
  arXiv:1612.03490 [astro-ph.CO].
  

  
\bibitem{Hazra:2016fkm} 
  D.~K.~Hazra, A.~Shafieloo, G.~F.~Smoot and A.~A.~Starobinsky,
  ``Primordial features and Planck polarization,''
  JCAP {\bf 1609}, no. 09, 009 (2016)
  [arXiv:1605.02106 [astro-ph.CO]].
  


\bibitem{Polarski:1995zn} 
  D.~Polarski and A.~A.~Starobinsky,
  ``Structure of primordial gravitational waves spectrum in a double inflationary model,''
  Phys.\ Lett.\ B {\bf 356}, 196 (1995)
  [astro-ph/9505125].
  
\bibitem{Lesgourgues:1998mq} 
  J.~Lesgourgues, D.~Polarski and A.~A.~Starobinsky,
  ``How large can be the primordial gravitational wave background in inflationary models?,''
  Mon.\ Not.\ Roy.\ Astron.\ Soc.\  {\bf 308}, 281 (1999)
  [astro-ph/9807019].
  
\bibitem{Polarski:1999fb} 
  D.~Polarski,
  ``Direct detection of primordial gravitational waves in a BSI inflationary model,''
  Phys.\ Lett.\ B {\bf 458}, 13 (1999)
  [gr-qc/9906075].
  
  
\bibitem{Chluba:2015bqa} 
  J.~Chluba, J.~Hamann and S.~P.~Patil,
  ``Features and New Physical Scales in Primordial Observables: Theory and Observation,''
  Int.\ J.\ Mod.\ Phys.\ D {\bf 24}, no. 10, 1530023 (2015)
  [arXiv:1505.01834 [astro-ph.CO]].
  





\bibitem{Xu:2016kwz} 
  Y.~Xu, J.~Hamann and X.~Chen,
  ``Precise measurements of inflationary features with 21 cm observations,''
  arXiv:1607.00817 [astro-ph.CO].

\bibitem{Chen:2016zuu} 
  X.~Chen, P.~D.~Meerburg and M.~M\"unchmeyer,
  ``The Future of Primordial Features with 21 cm Tomography,''
  arXiv:1605.09364 [astro-ph.CO].
  
  
\bibitem{Chen:2016vvw} 
  X.~Chen, C.~Dvorkin, Z.~Huang, M.~H.~Namjoo and L.~Verde,
  ``The Future of Primordial Features with Large-Scale Structure Surveys,''
  arXiv:1605.09365 [astro-ph.CO].
  
\bibitem{Ballardini:2016hpi} 
  M.~Ballardini, F.~Finelli, C.~Fedeli and L.~Moscardini,
  ``Probing primordial features with future galaxy surveys,''
  JCAP {\bf 1610}, 041 (2016)
  [arXiv:1606.03747 [astro-ph.CO]].







\bibitem{Cheung:2007st} 
  C.~Cheung, P.~Creminelli, A.~L.~Fitzpatrick, J.~Kaplan and L.~Senatore,
  ``The Effective Field Theory of Inflation,''
  JHEP {\bf 0803}, 014 (2008)
  [arXiv:0709.0293 [hep-th]].
  
  
\bibitem{Weinberg:2008hq} 
  S.~Weinberg,
  ``Effective Field Theory for Inflation,''
  Phys.\ Rev.\ D {\bf 77}, 123541 (2008)
  [arXiv:0804.4291 [hep-th]].




\bibitem{Achucarro:2014msa} 
  A.~Ach\'ucarro, V.~Atal, B.~Hu, P.~Ortiz and J.~Torrado,
  ``Inflation with moderately sharp features in the speed of sound: Generalized slow-roll and in-in formalism for power spectrum and bispectrum,''
  Phys.\ Rev.\ D {\bf 90}, no. 2, 023511 (2014)
  [arXiv:1404.7522 [astro-ph.CO]].
  
  
  
\bibitem{Achucarro:2013cva} 
  A.~Ach\'ucarro, V.~Atal, P.~Ortiz and J.~Torrado,
  ``Localized correlated features in the CMB power spectrum and primordial bispectrum from a transient reduction in the speed of sound,''
  Phys.\ Rev.\ D {\bf 89}, no. 10, 103006 (2014)
  [arXiv:1311.2552 [astro-ph.CO]].
  
  
  
  
\bibitem{Achucarro:2012fd} 
  A.~Ach\'ucarro, J.~O.~Gong, G.~A.~Palma and S.~P.~Patil,
  ``Correlating features in the primordial spectra,''
  Phys.\ Rev.\ D {\bf 87}, no. 12, 121301 (2013)
  [arXiv:1211.5619 [astro-ph.CO]].
  
\bibitem{Gong:2014spa} 
  J.~O.~Gong, K.~Schalm and G.~Shiu,
  ``Correlating correlation functions of primordial perturbations,''
  Phys.\ Rev.\ D {\bf 89}, no. 6, 063540 (2014)
  [arXiv:1401.4402 [astro-ph.CO]].
  
  
\bibitem{Fergusson:2014tza} 
  J.~R.~Fergusson, H.~F.~Gruetjen, E.~P.~S.~Shellard and B.~Wallisch,
  ``Polyspectra searches for sharp oscillatory features in cosmic microwave sky data,''
  Phys.\ Rev.\ D {\bf 91}, no. 12, 123506 (2015)
  [arXiv:1412.6152 [astro-ph.CO]].
  
  
\bibitem{Palma:2014hra} 
  G.~A.~Palma,
  ``Untangling features in the primordial spectra,''
  JCAP {\bf 1504}, no. 04, 035 (2015)
  [arXiv:1412.5615 [hep-th]].
  

\bibitem{Torrado:2016sls} 
  J.~Torrado, B.~Hu and A.~Ach\'ucarro,
  ``Robust predictions for an oscillatory bispectrum in Planck 2015 data from transient reductions in the speed of sound of the inflaton,''
  arXiv:1611.10350 [astro-ph.CO].
  
  
\bibitem{Mooij:2015cxa} 
  S.~Mooij, G.~A.~Palma, G.~Panotopoulos and A.~Soto,
  ``Consistency relations for sharp features in the primordial spectra,''
  JCAP {\bf 1510}, no. 10, 062 (2015)
  Erratum: [JCAP {\bf 1602}, no. 02, E01 (2016)]
  [arXiv:1507.08481 [astro-ph.CO]].
  
\bibitem{Cadavid:2015iya} 
  A.~G.~Cadavid, A.~E.~Romano and S.~Gariazzo,
  ``Effects of local features of the inflaton potential on the spectrum and bispectrum of primordial perturbations,''
  Eur.\ Phys.\ J.\ C {\bf 76}, no. 7, 385 (2016)
  [arXiv:1508.05687 [astro-ph.CO]].
  
\bibitem{Appleby:2015bpw} 
  S.~Appleby, J.~O.~Gong, D.~K.~Hazra, A.~Shafieloo and S.~Sypsas,
  ``Direct search for features in the primordial bispectrum,''
  Phys.\ Lett.\ B {\bf 760}, 297 (2016)
  [arXiv:1512.08977 [astro-ph.CO]].

\bibitem{Mooij:2016dsi} 
  S.~Mooij, G.~A.~Palma, G.~Panotopoulos and A.~Soto,
  ``Consistency relations for sharp inflationary non-Gaussian features,''
  JCAP {\bf 1609}, no. 09, 004 (2016)
  [arXiv:1604.03533 [astro-ph.CO]].



\bibitem{Meerburg:2015owa} 
  P.~D.~Meerburg, M.~M\"unchmeyer and B.~Wandelt,
  ``Joint resonant CMB power spectrum and bispectrum estimation,''
  Phys.\ Rev.\ D {\bf 93}, no. 4, 043536 (2016)
  [arXiv:1510.01756 [astro-ph.CO]].
  
  




 

\bibitem{Hu:2014hoa} 
  W.~Hu,
  ``Generalized slow-roll for tensor fluctuations,''
  Phys.\ Rev.\ D {\bf 89}, no. 12, 123503 (2014)
  [arXiv:1405.2020 [astro-ph.CO]].

\bibitem{Cai:2015yza} 
  Y.~Cai, Y.~T.~Wang and Y.~S.~Piao,
  ``Is there an effect of a nontrivial $c_T$ during inflation?,''
  Phys.\ Rev.\ D {\bf 93}, no. 6, 063005 (2016)
  [arXiv:1510.08716 [astro-ph.CO]].

  
\bibitem{Broy:2016zik} 
  B.~J.~Broy,
  ``Corrections to $n_s$ and $n_t$ from high scale physics,''
  Phys.\ Rev.\ D {\bf 94}, no. 10, 103508 (2016)
  Addendum: [Phys.\ Rev.\ D {\bf 94}, no. 10, 109901 (2016)]
  [arXiv:1609.03570 [hep-th]].
  
\bibitem{Maldacena:2002vr} 
  J.~M.~Maldacena,
  ``Non-Gaussian features of primordial fluctuations in single field inflationary models,''
  JHEP {\bf 0305}, 013 (2003)
  [astro-ph/0210603].

  
  
\bibitem{Weinberg:2005vy} 
  S.~Weinberg,
  ``Quantum contributions to cosmological correlations,''
  Phys.\ Rev.\ D {\bf 72}, 043514 (2005)
  [hep-th/0506236].
  
  
\bibitem{Stewart:2001cd} 
  E.~D.~Stewart,
  ``The Spectrum of density perturbations produced during inflation to leading order in a general slow-roll approximation,''
  Phys.\ Rev.\ D {\bf 65}, 103508 (2002)
  [astro-ph/0110322].
  
  
\bibitem{Choe:2004zg} 
  J.~Choe, J.~O.~Gong and E.~D.~Stewart,
  ``Second order general slow-roll power spectrum,''
  JCAP {\bf 0407}, 012 (2004)
  [hep-ph/0405155].
  
\bibitem{Adshead:2011bw} 
  P.~Adshead, W.~Hu, C.~Dvorkin and H.~V.~Peiris,
  ``Fast Computation of Bispectrum Features with Generalized slow-roll,''
  Phys.\ Rev.\ D {\bf 84}, 043519 (2011)
  [arXiv:1102.3435 [astro-ph.CO]].
  
\bibitem{Dvorkin:2009ne} 
  C.~Dvorkin and W.~Hu,
  ``Generalized slow-roll for Large Power Spectrum Features,''
  Phys.\ Rev.\ D {\bf 81}, 023518 (2010)
  [arXiv:0910.2237 [astro-ph.CO]].
  

  
\bibitem{Baumann:2009ds} 
  D.~Baumann,
 ``Inflation,''
  arXiv:0907.5424 [hep-th].
  
  
  
 
  
\bibitem{Achucarro:2010jv} 
  A.~Ach\'ucarro, J.~O.~Gong, S.~Hardeman, G.~A.~Palma and S.~P.~Patil,
  ``Mass hierarchies and non-decoupling in multi-scalar field dynamics,''
  Phys.\ Rev.\ D {\bf 84}, 043502 (2011)
  [arXiv:1005.3848 [hep-th]].
  
  
\bibitem{Achucarro:2010da} 
  A.~Ach\'ucarro, J.~O.~Gong, S.~Hardeman, G.~A.~Palma and S.~P.~Patil,
  ``Features of heavy physics in the CMB power spectrum,''
  JCAP {\bf 1101}, 030 (2011)
  [arXiv:1010.3693 [hep-ph]].
  
  
\bibitem{Tolley:2009fg} 
  A.~J.~Tolley and M.~Wyman,
  ``The Gelaton Scenario: Equilateral non-Gaussianity from multi-field dynamics,''
  Phys.\ Rev.\ D {\bf 81}, 043502 (2010)
  [arXiv:0910.1853 [hep-th]].
  
  
  
  
  
 
  
 

  
\bibitem{Flauger:2009ab} 
  R.~Flauger, L.~McAllister, E.~Pajer, A.~Westphal and G.~Xu,
  ``Oscillations in the CMB from Axion Monodromy Inflation,''
  JCAP {\bf 1006}, 009 (2010)
  [arXiv:0907.2916 [hep-th]].


 
\bibitem{Freese:1990rb} 
  K.~Freese, J.~A.~Frieman and A.~V.~Olinto,
  ``Natural inflation with pseudo - Nambu-Goldstone bosons,''
  Phys.\ Rev.\ Lett.\  {\bf 65}, 3233 (1990).
  
  
\bibitem{Lyth:1996im} 
  D.~H.~Lyth,
  ``What would we learn by detecting a gravitational wave signal in the cosmic microwave background anisotropy?,''
  Phys.\ Rev.\ Lett.\  {\bf 78}, 1861 (1997)
  [hep-ph/9606387].

\bibitem{Easther:2006qu} 
  R.~Easther, W.~H.~Kinney and B.~A.~Powell,
  ``The Lyth bound and the end of inflation,''
  JCAP {\bf 0608}, 004 (2006)
  [astro-ph/0601276].


\bibitem{Ade:2015lrj} 
  P.~A.~R.~Ade {\it et al.} [Planck Collaboration],
  ``Planck 2015 results. XX. Constraints on inflation,''
  Astron.\ Astrophys.\  {\bf 594}, A20 (2016)
  [arXiv:1502.02114 [astro-ph.CO]].
  
  
\bibitem{Obata:2016xcr} 
  I.~Obata and J.~Soda,
  ``Oscillating Chiral Tensor Spectrum from Axionic Inflation,''
  Phys.\ Rev.\ D {\bf 94}, no. 4, 044062 (2016)
  [arXiv:1607.01847 [astro-ph.CO]].


\bibitem{Creminelli:2014wna} 
  P.~Creminelli, J.~Gleyzes, J.~Nore\~na and F.~Vernizzi,
 ``Resilience of the standard predictions for primordial tensor modes,''
  Phys.\ Rev.\ Lett.\  {\bf 113}, no. 23, 231301 (2014)
  [arXiv:1407.8439 [astro-ph.CO]].
  
  
  
\end{thebibliography}
\end{document}